\documentclass[12pt]{iopart}
\usepackage{blindtext}
\pdfoutput=1
\usepackage[utf8]{inputenc}
\expandafter\let\csname equation*\endcsname\relax
\expandafter\let\csname endequation*\endcsname\relax
\usepackage{amsmath} 
\usepackage{graphicx}
\usepackage{amssymb}
\usepackage{hyperref}
\usepackage{amsthm}
\usepackage{csquotes}
\usepackage{textcomp, gensymb}
\usepackage{float} 
\graphicspath{ {./images/} }
\usepackage{caption}
\usepackage{appendix}
\usepackage{xcolor}
\usepackage{subcaption}
\usepackage[T1,T2A]{fontenc}
\usepackage[ukrainian, greek, english]{babel}
\usepackage[or]{teubner}
\usepackage{soul}
\usepackage{gensymb}

\usepackage{iopams}  
    \usepackage[
    backend=bibtex,
   style=nature,
   doi=false, url=false,labeldate=year,date=year,
  ]{biblatex}

\DeclareUnicodeCharacter{2212}{-}
\DeclareUnicodeCharacter{0301}{\hspace{-1ex}\'{ }}

\begin{document}

\title[]{Critical and tricritical singularities from small-scale Monte Carlo simulations: The Blume-Capel model in two dimensions}

\author{Leïla~Moueddene$^{1,2,3}$, Nikolaos~G~Fytas$^4$, Yurij~Holovatch$^{5,2,3,6}$, Ralph~Kenna$^{2,3}$, Bertrand~Berche$^{1,3}$ }
\ead{leila.moueddene@univ-lorraine.fr}

\address{$^1$ Laboratoire de Physique et Chimie Théoriques, CNRS - Université de Lorraine, UMR 7019, Nancy, France}
\address{$^2$ Centre for Fluid and Complex Systems, Coventry University, Coventry CV1 5FB, United Kingdom}
\address{$^3$ $\mathbb{L}^4$ Collaboration \& Doctoral College for the Statistical Physics of Complex Systems, Leipzig-Lorraine-Lviv-Coventry}
\address{$^4$ School of Mathematics, Statistics and Actuarial Science, University of Essex, Colchester CO4 3SQ, United Kingdom}
\address{$^5$ Institute for Condensed Matter Physics, National Acad. Sci. of Ukraine, 79011 Lviv, Ukraine}
\address{$^6$  Complexity Science Hub Vienna, 1080 Vienna, Austria}

\begin{abstract}
We show that the study of critical properties of the Blume-Capel model at two dimensions can be deduced from Monte Carlo simulations with good accuracy even for small system sizes when one analyses the behaviour of the zeros of the partition function. The phase diagram of the model displays a line of second-order phase transitions ending at a tricritical point, then a line of first-order transitions. We concentrate on critical and tricritical properties and compare the accuracy achieved via standard finite-size scaling of thermodynamic quantities with that from the zeros analysis. This latter analysis showcases spectacular precision, even for systems as small as $64$ spins! We also show that the zeros are very sensitive to subtle crossover effects.
\end{abstract}

\submitto{Journal of Statistical Mechanics}

\maketitle

\emph{Ralph Kenna passed away on October 26th, 2023, while preparing this manuscript. He was an excellent friend, a stimulating collaborator, a tireless mentor. But for us, he was much more than that and he will be missed so much. This and many other works have been initiated at his contact. He will not have the opportunity to put the final touches on this work which is then much weaker than it would have been, impregnated from Ralph's spirit}.

\section{Introduction}
\label{sec:intro}

Global warming is nowadays an essential concern for everyone, and the indigence of the most powerful governments which do not take measures to meet the challenges leads more and more people, the youngest in particular, to seize the environmental issues. The message of science is unequivocal: there is no collective solution without a considerable reduction in greenhouse gas emissions. Academic circles play their part in this awareness, some in a very active manner, such as \href{https://scientistrebellion.org}{Scientists Rebellion}. At universities, researchers wonder about the compatibility between their professional practices and reducing the carbon footprint. In the field of intensive computing, supercomputing centres enormously generate greenhouse gases and perhaps the use of (massively parallel) large-scale simulations and high-performance computing must be questioned. We consider here an approach which goes into the opposite direction of \emph{small-scale simulations} in the field of critical phenomena and demonstrate that it is possible to obtain relatively precise results at a significantly lower computational cost. In particular, we focus on the determination of critical exponents and other universal amplitudes of a well-known lattice model in statistical physics which features tricritical behaviour.

Thermodynamic quantities in the vicinity of a second-order phase transition are indeed characterized by power laws in the reduced temperature $t$ and external field $h$. In the thermodynamic limit, at $h=0$, the divergence of the correlation length is responsible for the critical singularities, following from $ \xi_{\infty} (t) \sim |t|^{-\nu}$. Additionally, one has $C_{\infty} (t) \sim |t|^{-\alpha}$ for the specific heat, $\chi_{\infty} (t) \sim |t|^{-\gamma}$ for the susceptibility, and  $m_{\infty} (t) \sim |t|^{\beta}$ for the order parameter or magnetisation of the system at $T < T_{\rm c}$. The critical behaviour may be identical for distinct systems, therefore the set of exponents determine a universality class. These universality classes depend only on very general properties, like the dimensionality of the system, the range of interactions, and the symmetries of the ground state of the system.

Our aim here is to show that the analysis of the scaling of the zeros of the partition function, popularized by the seminal work of Lee and Yang~\cite{Yang,Lee}, is a very accurate tool for extracting the critical behaviour of spin models. This is of course not new, but we move here in the opposite direction than the usual trend. Ordinarily, critical exponents are extracted from extensive Monte Carlo simulations via finite-size scaling (FSS) methods of thermodynamic quantities like the susceptibility or the order parameter. This requires simulations at different sizes for a range of temperatures to determine first the pseudocritical parameters where the diverging quantities (\emph{e.g.}, the susceptibility) display their maximal values, then to analyze the size-dependent power-law behaviours of these thermodynamic observables at pseudocriticality. The larger the sizes, the better the accuracy of the critical exponents measured, especially in cases where corrections to scaling are strong and call for particular attention~\cite{Fytas_PRL}.

The zeros of the partition function on the other hand can provide an alternative to these commonly used thermodynamic quantities. In fact, their behaviour upon approaching the critical point also encodes the critical exponents needed to characterize a universality class. We show here, by comparing the two approaches outlined above, that, indeed, the FSS analysis of the partition function zeros does not require as large system sizes as those needed in the case where the usual thermodynamic quantities are involved in order to achieve a given accuracy for the estimation of the basic critical exponents. In this context, our work designates the following unexpected finding: An excellent precision can be achieved via scrutinizing the partition function zeros of very small system sizes, of the order of $L\le 8$, where $L$ denotes the linear dimension of the lattice; thus, as few as $64$ spins or less for a two-dimensional (2D) lattice considered here. To support our claim we choose as a platform the two-dimensional Blume-Capel model~\cite{CAPEL, Blume}, one of the most well-studied fruit-fly models of statistical physics, which has a rich phase diagram featuring first- and second-order phase transition lines that merge at a tricritical point. Of course, the numerical results presented here do not consist a rigorous proof of our main conclusion, but provide a clear and strong empirical support to it, an aspect that has not been recorded previously in the relevant literature. As a side remark, the analysis of the zeros that we provide is to the best of our knowledge the most accurate ever performed for the Blume-Capel model.

The rest of the paper is laid out as follows: In section~\ref{sec:BC_model} we introduce the model and in section~\ref{sec:numerics} we shortly outline the implemented numerical methods. Our numerical results and FSS analysis are presented in sections~\ref{sec:FSS_standard}, \ref{sec:zeros}, and \ref{sec:cumulants}. Finally, a summary of our conclusions is given in section~\ref{sec:conclusions}.

\section{Blume-Capel model}
\label{sec:BC_model}

The Blume-Capel model is a particular case of the Blume-Emery-Griffiths model~\cite{CAPEL, Blume} which has greatly contributed to the study of tricriticality in condensed-matter physics. It is also a model of great experimental interest since it can describe many physical systems, including, among others, ultracold quantum gases~\cite{Shin}, $^3$He-$^4$He mixtures, and multi-component fluids~\cite{PhysRevA.4.1071,Lawrie}. In fact, experimental studies have focused on systems with a tricritical point, such as FeCl$_2$ for example~\cite{FeCl2Shang}. 
The spin-$1$ Blume-Capel model is introduced through the Hamiltonian 
\begin{equation}
    \mathcal{H}=-J\sum_{\langle i,j\rangle}\sigma_i \sigma_j + \Delta \sum_i \sigma_i^2 - H\sum_i \sigma_i =  E_J + \Delta E_{\Delta}-HM,
\end{equation}
where $J > 0$ denotes the ferromagnetic exchange interaction coupling, the spin variables $\sigma_i \in \{−1, 0, +1\}$ live on a square lattice with periodic boundaries, and $\langle i,j \rangle$ indicates summation over nearest neighbors. 
$\Delta$ represents the crystal-field strength that controls the density of vacancies $(\sigma_i = 0)$. For $\Delta = -\infty$ vacancies are suppressed and the Hamiltonian reduces to that of the Ising ferromagnet.

The Blume-Capel model has been studied extensively, mainly using numerical approaches~\cite{PhysRevB.33.1717,PhysRevE.73.036702,malakis10,Kwak2015,butera,PhysRevE.102.062138}; for a recent review, we refer the reader to
reference~\cite{Zierenberg}. 
\begin{figure}[H]
\centering
\includegraphics[width=0.7\textwidth]{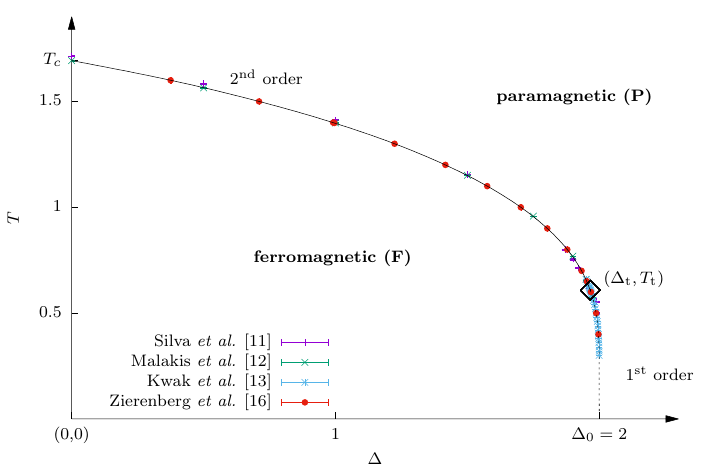}
\caption{Phase diagram of the 
square-lattice Blume-Capel model in the ($\Delta$, $T$)--plane showing the ferromagnetic (\textbf{F}) and paramagnetic (\textbf{P}) phases that are separated by a continuous transition at small $\Delta$ (solid line) and a first-order at large $\Delta$ (dotted line). The line segments meet at a tricritical point $(\Delta_{\rm t}, \; T_{\rm t})$ marked by a black rhombus. Numerical data shown are selected estimates from previous studies, including the critical point at zero crystal field which we denote hereafter for the needs of our work as $T_{\rm c} (\Delta = 0)$, or simply as $T_{\rm c}$. The numerical simulations reported in this paper were carried out at these two special points $(\Delta, \; T) = \{(0, \; T_{\rm c}), \;(\Delta_{\rm t}, \; T_{\rm t})\}$.}
\label{fig:phase_diagram}
\end{figure}
A reproduction of the phase diagram of the square-lattice Blume-Capel model in the ($\Delta$, $T$)--plane is shown
in figure~\ref{fig:phase_diagram}: For small $\Delta$ there is a line of continuous transitions between the ferromagnetic and paramagnetic phases that crosses the vertical axis at $(\Delta_{\rm c} = 0, \; T_{\rm c}\approx 1.6929)$~\cite{malakis10}. For large $\Delta$, on the other hand, the transition becomes discontinuous and it meets the $T=0$ line at $\Delta_0 = zJ/2$~\cite{CAPEL}, where $z = 4$ is the coordination number (here we set $J = 1$ and $k_{\rm B} = 1$ to fix the energy and temperature scale). The two line
segments meet in a tricritical point estimated to be at
$(\Delta_{\rm t}\approx1.966, \; T_{\rm t}\approx0.608)$~\cite{Kwak2015,Jung}. It is
well established that the second-order transitions belong to the universality class of the two-dimensional Ising model~\cite{Zierenberg,fytas11}.
We should note here that similar phase diagrams showcasing a tricritical point can also be found in the Baxter-Wu model in transverse magnetic field~\cite{Capponi_2014}, the spin-$1$ Baxter-Wu model~\cite{fytas21,vasilopoulos22,macedo23} and the bond-diluted 4-state Potts model at three dimensions~\cite{chatelain1,chatelain2}. Note that the two-dimensional disordered $4$-state Potts model always displays a continuous phase transition~\cite{holovatch2004,monthus}.

\section{Numerical methods}
\label{sec:numerics}

\subsection{Hybrid and Wang-Landau simulations}

Most conventional Monte Carlo schemes generate a canonical distribution $P(E,\beta)=g(E)e^{-\beta E}$ at a fixed temperature $\beta$ (to be precise $\beta$ refers to the inverse temperature and should not be confused with the critical exponent of the order parameter). Then, typically as many simulations as necessary are performed in order to probe the thermodynamic observables under study within a sufficient temperature range. For the particular case of the Blume-Capel model and due to the presence of the zero-state spins, we employ here a combination of a Wolff single-cluster update~\cite{wolff} of the $\pm 1$ spins and a single-spin-flip Metropolis update~\cite{hybrid,hasenbusch10,malakis12} to account for the vacancies, denoted herewith as the hybrid Metropolis-Wolff (MW) approach.

We also performed an additional array of canonical Monte Carlo simulations implementing the Wang-Landau (WL) entropic sampling method~\cite{PhysRevLett.86.2050,PhysRevE.64.056101, landau_wang_2004}, an efficient protocol that samples the density of states $g(E)$. As $g(E)$ does not depend on $\beta$ (or $\Delta$), it is possible to construct the canonical distribution at any temperature and then compute the partition function. Using the Wang-Landau method the energy and magnetisation (in zero magnetic field) are computed as follows
\begin{equation}
   \langle E\rangle =\frac{\sum_{E_J,E_\Delta} (-JE_J+\Delta E_\Delta)g(E_J,E_\Delta) e^{-\beta (-JE_J+\Delta E_\Delta)}}{\sum_{E_J,E_\Delta} g(E_J,E_\Delta) e^{-\beta (-JE_J+\Delta E_\Delta)}},
\end{equation}
\begin{equation}
   \langle|M|\rangle =\frac{\sum_{E_J,E_\Delta,M} |M| g(E_J,E_\Delta,M) e^{-\beta (-JE_J+\Delta E_\Delta)}}{\sum_{E_J,E_\Delta,M} g(E_J,E_\Delta,M) e^{-\beta (-JE_J+\Delta E_\Delta)}},
\end{equation}
where $g(E_J,E_\Delta)$ and $g(E_J,E_\Delta,M)$ are the appropriate densities of states and the absolute value of $M$ breaks the $Z_2$ symmetry as usual (otherwise the simulations would deliver a vanishing magnetisation in zero magnetic field).

We should point out at this stage that the main drawback of the Wang-Landau algorithm lies in its heavy computational time cost which limits simulations to rather moderate system sizes. Of course, several improvements of the algorithm have been presented over the last years facilitating the simulations of much larger systems~\cite{zhou03,malakis04,belardinelli07,netto08,vogel13}, some of which however with unclear consequences in the numerical accuracy~\cite{fytas08}. In lieu, the hybrid approach~\cite{hybrid,hasenbusch10,malakis12} allows the simulation of considerably larger system sizes in moderate computational times but lacks direct access to the density of states. However, this problem can be surpassed with the help of histogram reweighting~\cite{PhysRevLett.61.2635}. This is a well-known technique that allows the safe extrapolation of data emerging from a Monte Carlo simulation at a fixed temperature to a nearby temperature window (within a certain level of accuracy).

For both methods (hybrid and Wang-Landau), the specific heat ($C$), susceptibility ($\chi$), magnetocaloric ($m_{T}$), and ``mixed'' ($\chi_{12}$ and $\chi_2$) coefficients (see also below) are accessible through the usual fluctuation-dissipation relations
\begin{equation}
C = \frac{\langle E^{2}\rangle -\langle E\rangle ^{2}}{k_{\rm B}T^2},
\end{equation}
\begin{equation}
\chi = \frac{\langle M^2\rangle -\langle |M|\rangle ^2}{k_{\rm B}T},
\end{equation}
\begin{equation}
m_{T} = |{\langle E|M|\rangle -\langle E\rangle \langle |M|\rangle }|,\label{eq-m_T}
\end{equation}
\begin{equation}
\chi_{12} = |{\langle E_\Delta|M|\rangle -\langle E_\Delta\rangle \langle |M|\rangle }|,
\label{eq-chi12}
\end{equation}
and 
\begin{equation}
\chi_2 = \frac{\langle E_\Delta^2\rangle -\langle |E_\Delta|\rangle ^2}{k_{\rm B}T}.
\label{eq-chi2}
\end{equation}
Note that in the above formulas, the notation $E$ refers to the total energy, whereas $E_\Delta$ when only the contribution of the crystal-field term is taken into account [equations~(\ref{eq-chi12}) and (\ref{eq-chi2})].

As already mentioned above, we carried out several series of simulations for the square-lattice Blume-Capel model with periodic boundary conditions. In particular, for the hybrid method we run simulations at the zero crystal-field critical point~\cite{malakis10} and the tricritical point~\cite{Kwak2015}, located along the phase boundary of figure~\ref{fig:phase_diagram} at
\begin{equation}
(\Delta_{\rm c},\; T_{\rm c}) = (0, \; 1.6929)\;; \;\; (\Delta_{\rm t},\; T_{\rm t}) = (1.966, \; 0.608),
\label{eq-points}
\end{equation}
respectively, establishing the targeted simulation points of our numerical endeavour. Our results were then extrapolated to nearby temperature values benefiting from the histogram reweighting technique. In our protocol the hybrid algorithm performs $1500 \times N$ Monte Carlo Steps per spin (MCSs) to reach a steady state and $9 \times 1500 \times N$ MCSs to collect the data; $N = L \times L$ denotes the total number of spins on the lattice. A MCS consists of $3N$ Metropolis steps followed by $L$ Wolff steps, a practice suggested as optimum in references~\cite{Zierenberg,fytas18}. For the Wang-Landau method now, the number of iterations increased until the modification factor $f$, which controls the convergence of the simulation, was less than $f_{\rm final}=1+10^{-8}$~\cite{PhysRevE.64.056101}. 
\begin{figure}[H]
\centering
\includegraphics[width=0.5\textwidth]{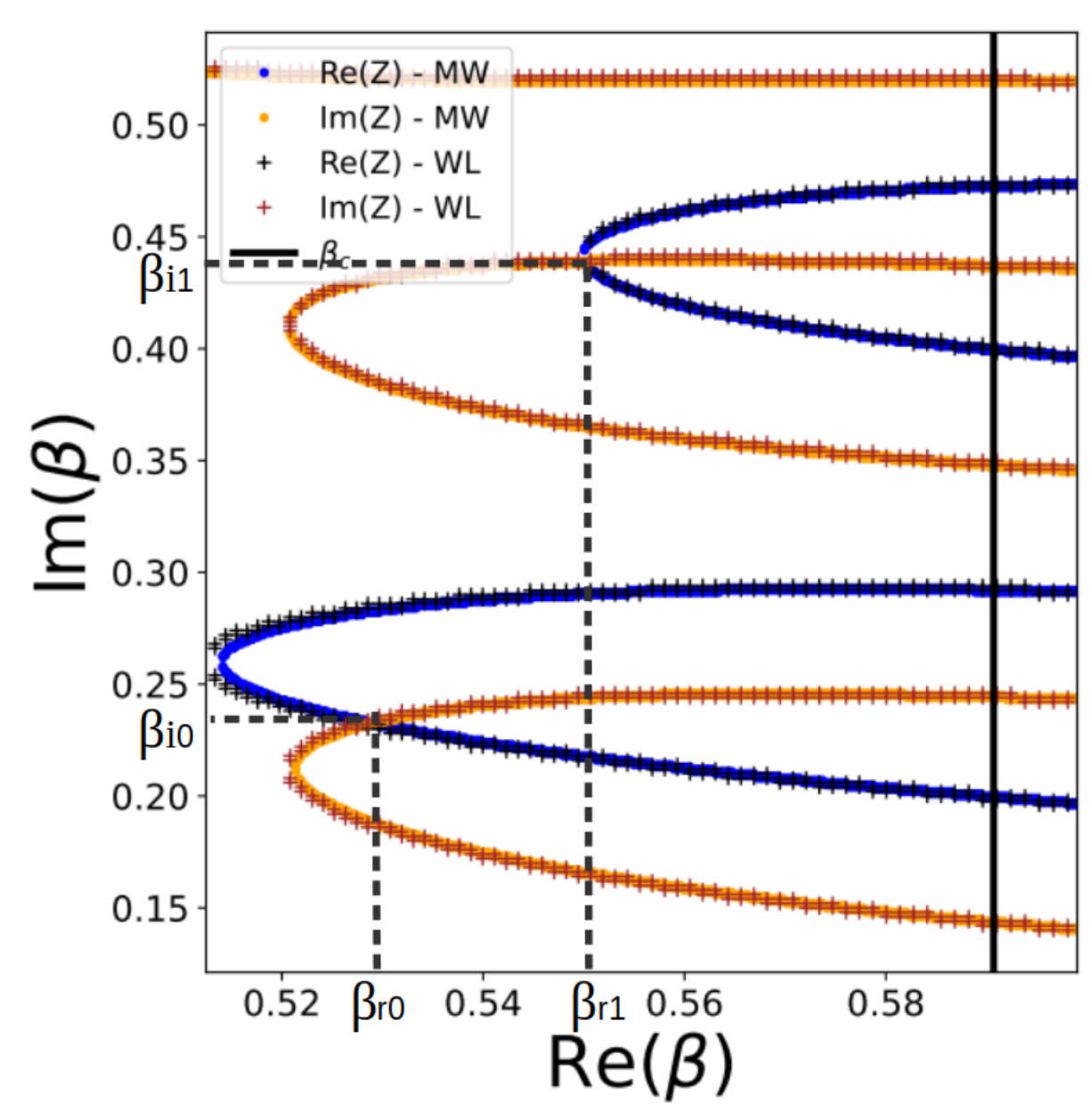}
\caption{Fisher zeros at the zero crystal-field critical point specified
in equation (9) in the complex temperature plane for $L = 4$. The two sets of blue and yellow circles show respectively changes in the sign of the real and imaginary parts of the partition function obtained from hybrid Metropolis-Wolff algorithm, while the sets of black and red crosses those obtained from Wang-Landau simulations. A crossing of the lines is a zero of the partition function. The first two Fisher zeros are denoted as $[\beta_{r1}, \beta_{i1}]$  and  $[\beta_{r2}, \beta_{i2}]$.}\label{fig:fisher_zeros}
\end{figure}
Finally, errors were determined using the Jackknife method~\cite{efronjackknife} for the hybrid algorithm, while $10$ simulations with different initial configurations were carried out to compute errors of the relevant thermodynamic quantities accumulated during the Wang-Landau simulations. 

\subsection{Zeros of the partition function}

The partition function of the Blume-Capel model on a finite $D$-dimensional lattice
may be given in terms of the density of states $g(E_J,E_\Delta,M)$
\begin{equation}
    \mathcal{Z}(\beta)=\sum_{E_J,E_\Delta,M} g(E_J,E_\Delta,M) e^{-\beta (-JE_J+\Delta E_\Delta-HM)}.
\end{equation}
The microstates are labelled by the energies $E_J$ (integer values $\{-DN, DN\}$),  the crystal-field energies $E_\Delta$ (positive integer values $\{0,N\}$), and the magnetisation values (integer values $\{-N, N\}$). The sum  is explicitly written as
 \begin{equation}
\mathcal{Z}_L= \sum_{E_J=-DN}^{DN} \sum_{E_\Delta=0}^{N} \sum_{M=-N}^{N}  g(E_J,E_\Delta,M) e^{-\beta E - \delta E_\Delta + hM },
\end{equation}
where $h=\beta H$ and $\delta=\beta \Delta$ --  we stress that there should be no confusion with the critical exponent $\delta$ here, as the inverse temperature should not be confused with the critical exponent $\beta$.

In order to find Fisher zeros -- which are the zeros of the partition function in the complex plane of the (inverse) temperature variable $\beta = \beta_r+i\beta_i$ -- the partition function in zero magnetic field  takes the form 
\begin{equation}
\mathcal{Z}(\beta_r,\beta_i)=\sum_{E_J,E_\Delta} g(E_J,E_\Delta) e^{-\beta_r(-JE_J+\Delta E_\Delta)} (\cos[\beta_i (JE_J-\Delta E_\Delta)]+i\sin[\beta_i (JE_J-\Delta E_\Delta)])\label{EQ9}
\end{equation}
where $g(E_J,E_\Delta)$ is the associated density of states.  
One defines a normalised version of equation~(\ref{EQ9}) by $Z(\beta_r,\beta_i)=\mathcal{Z}(\beta)/\mathcal{Z}[\operatorname{Re}(\beta)]$, which selects only the sine and cosine parts 
\begin{align} 
Z(\beta_r,\beta_i) & =\frac{\sum_{E_J,E_\Delta} g(E_J,E_\Delta) e^{-\beta_r(-JE_J+\Delta E_\Delta)} (\cos[\beta_i (JE_J-\Delta E_\Delta)]+i\sin[\beta_i (JE_J-\Delta E_\Delta)])}{\sum_{E_J,E_\Delta} g(E_J,E_\Delta) e^{-\beta_r(-JE_J+\Delta E_\Delta)}} \nonumber \\
 & = \langle\cos[\beta_i (JE_J-\Delta E_\Delta)]\rangle_{\beta_r} + i\langle \sin[\beta_i (JE_J-\Delta E_\Delta)]\rangle_{\beta_r}.
\label{eq-split}
\end{align}

The zeros of the partition function, labelled by the index $n$, are the values $\beta_{rn}+i\beta_{in}$ of the complex temperature where both real and imaginary parts of $Z(\beta_r,\beta_i)$ are equal to zero. 
In order to find the first zero $\beta_{r0}+i\beta_{i0}$, simulations were performed in a range of complex temperatures $(\beta_r,\beta_i)$ in the vicinity of the real values of the parameters at the transition (\emph{i.e.}, on the transition line) to get the intersection in the complex plane between two curves: Those corresponding to the independent vanishing of the real part $\langle\cos[\beta_i (JE_J-\Delta E_\Delta)]\rangle_{\beta_r} = 0$ on one hand and of the imaginary part $\langle\sin[\beta_i (JE_J-\Delta E_\Delta)]\rangle_{\beta_r}=0$ on the other hand. 
When both conditions are met a zero is found. The closest to the transition line is the first zero. At the tricritical point -- see equation~(\ref{eq-points}) -- and for sizes $L=4 - 8$, the zeros were found using the Wang-Landau algorithm working exactly at $\Delta_{\rm t} = 1.966$. For the larger sizes considered within the range $L = 16 - 64$, the simulations were performed with the hybrid algorithm enhanced by the histogram reweighting technique to explore the phase space around the simulated temperature $T_{\rm t}$ with the minimal effort. An analogous strategy was also followed at the zero crystal field critical point.

\begin{figure}[H]
    \centering
\hbox{     \begin{subfigure}[b]{0.5\textwidth}
        \includegraphics[width=\textwidth]{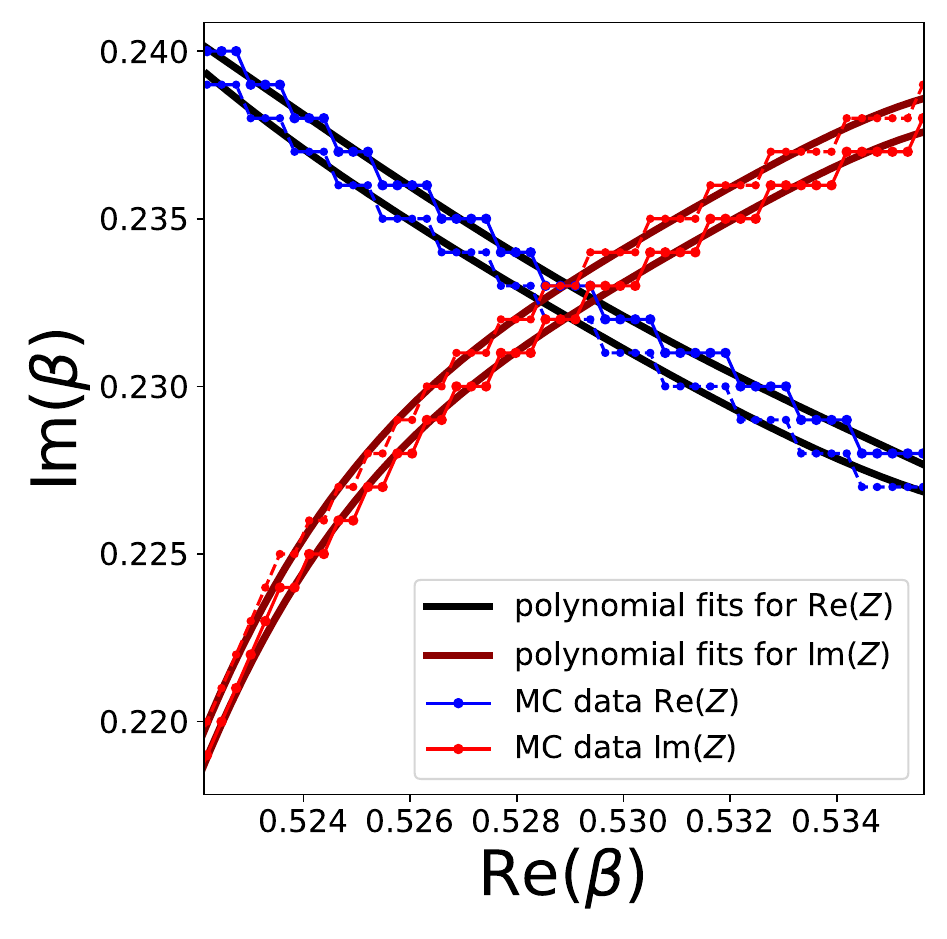}
    \end{subfigure}
        \begin{subfigure}[b]{0.5\textwidth}
        \includegraphics[width=\textwidth]{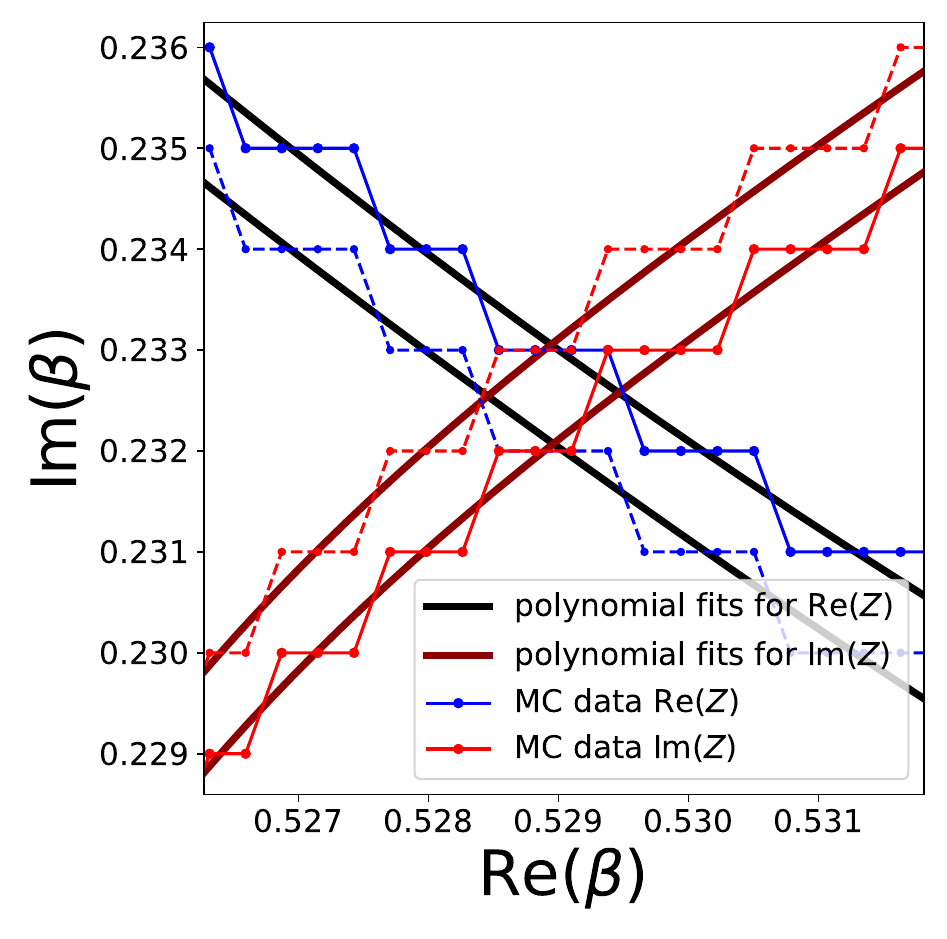}
    \end{subfigure}
    }
\caption{\textbf{Left panel}: Fisher zeros at $\Delta_{\rm c} = 0$ for a system with linear size $L=4$. A single hybrid Metropolis-Wolff simulation was performed at the critical temperature $T_{\rm c} = 1.6929$. The two sets of blue and red circles show respectively changes in the sign of the real and imaginary parts of the partition function. The solid lines are fourth-order degree polynomial fits and the zero sought is enclosed within the central losange. \textbf{Right panel}: A zoom in of the interesting area for the benefit of the reader.}\label{fig:fisher_zeros_zoom}
\end{figure}

A consistency check regarding the validity of the method and the numerical accuracy between the different algorithms is provided in figure~\ref{fig:fisher_zeros}. This plot depicts the curves in the complex temperature plane at $\Delta_{\rm c} = 0$, where either the real or the imaginary parts of $Z(\beta_r,\beta_i)$ vanish. Comparative results from hybrid Metropolis-Wolff and Wang-Landau methods are shown for a system with linear size $L=4$. It is evident that there is a perfect matching between the numerical data produced by the two algorithms. The first zeros, given by the intersections of the two types of curves, are located in the vicinity of the critical temperature (shown by the vertical solid line). In order to spot an accurate location of the zeros, we follow a two-step process: During the first step the approximate position of the zeros is graphically identified by plotting the sign changes of $\operatorname{Re} (Z)$ and $\operatorname{Im} (Z)$ in the vicinity of the critical (or tricritical) temperature. In the second step and once these positions are well approximated, a polynomial fit over a few points in the vicinity of the intersection using the standard Newton algorithm determines with high precision where the real and imaginary parts of the function are simultaneously equal to zero; see figure~\ref{fig:fisher_zeros_zoom} for a graphical illustration of the described approach.  

\subsection{Some comments on the goodness of fits}

\begin{figure}[ht]
\centering
\includegraphics[width=1\textwidth]{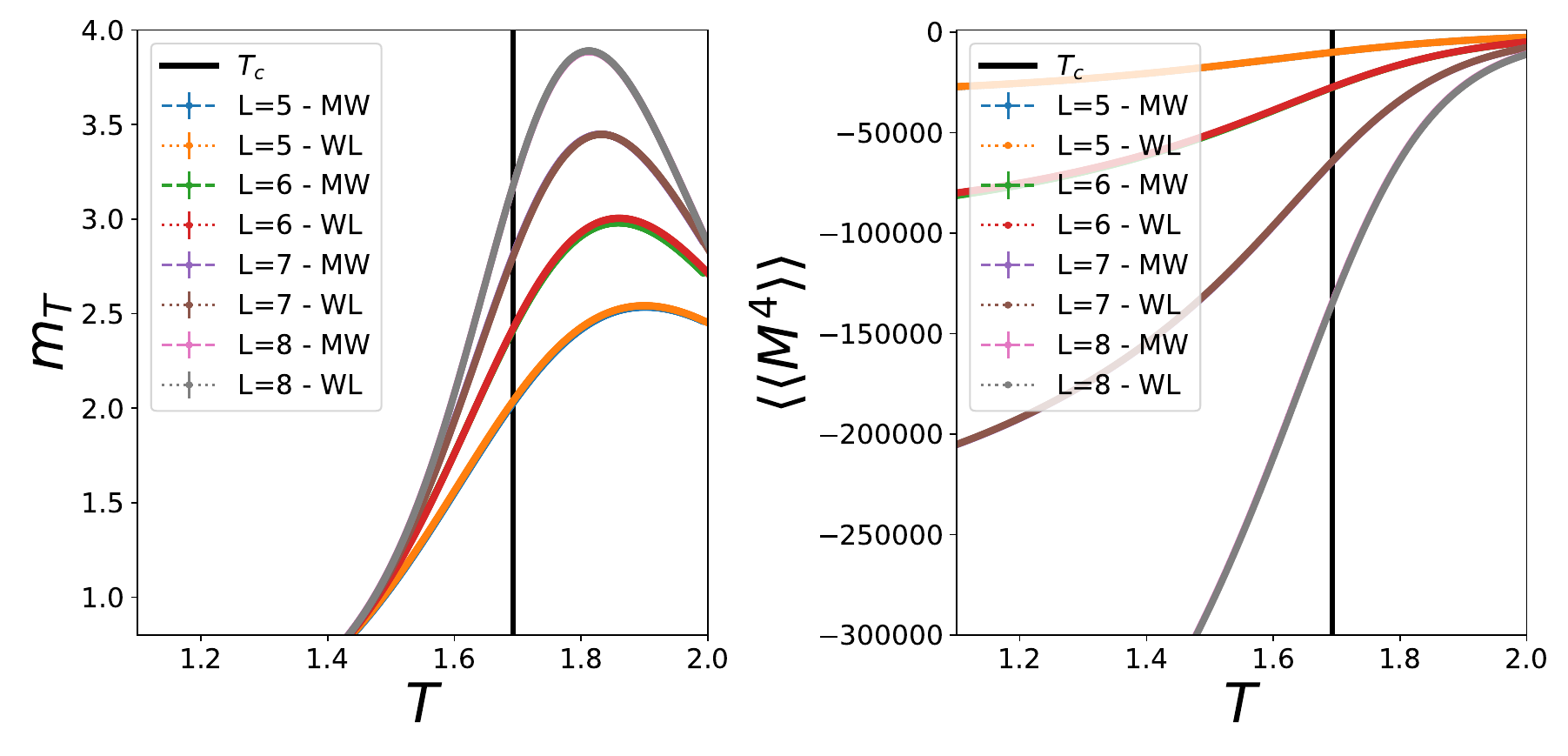}
\caption{Magnetocaloric coefficient (\textbf{left panel}) and fourth-order magentisation cumulant (\textbf{right panel}) vs. temperature. Comparison of numerical data in the small lattice-size regime at $\Delta = 0$ between hybrid Metropolis-Wolff  and Wang-Landau simulations.}\label{fig:comparison}
\end{figure}

For the fitting procedure discussed below and in order to determine an acceptable fit we employed the standard $\chi^2$ test for goodness of fit -- this $\chi^2$ parameter here should not be confused with the observable $\chi_2$! Specifically, the $Q$ value of our $\chi^2$ test is the probability of finding a $\chi^2$ value which is even larger than the one actually found
from our data~\cite{young}. Recall that this probability is computed by
assuming (1) Gaussian statistics and (2) the correctness of
the fit’s functional form. We consider a fit as being fair only
if $10\% < Q < 90\%$. Note that if $Q < 0.001$ the model should be well verified to find the cause but can still be acceptable~\cite{NumericalRecipes}. 

\section{Finite-size scaling for standard thermodynamic quantities}
\label{sec:FSS_standard}

\subsection{Scaling hypothesis and finite-size scaling at second-order phase transitions}

At the critical point, for systems of finite size $L$, the singular part of the free energy density is a generalized homogeneous function~\cite{PatashinskiiProkowskiiBook}, expressed according to the scaling fields $t = T-T_{\rm c}(\Delta_{\rm c} = 0)$ and $h = H$ 
\begin{equation}
    f_{\rm sing}(t,h)=L^{-D}f(L^{y_t}t, L^{y_h}h)
    \label{eq-Homf}
\end{equation}
with $y_t$ and $y_h$ the renormalization-group  eigenvalues associated to $t$ and $h$. 
First-order derivatives with respect to the temperature $t$ and magnetic field $h$ define the internal energy $e$ and magnetisation, respectively, while second-order derivatives provide the specific heat $C$, susceptibility $\chi$, and magnetocaloric coefficient  $m_{T}$ [cross derivative, $m_{T} = \partial^2f_{\rm sing}(t,h)/\partial t\partial h$; see equation~(\ref{eq-m_T})].  Homogeneous forms analogous to equation~(\ref{eq-Homf}) 
also follow for these quantities, leading at the critical point $t=0$, $h=0$ to the following FSS behaviour
\begin{equation}
    e_L\sim L^{-D+y_t},\quad c_L\sim L^{-D+2y_t},\quad m_L\sim L^{-D+y_h},\quad \chi_L\sim L^{-D+2y_h},\quad m_{T_{L}}\sim L^{-D+y_t+y_h}.
\end{equation}
Note the renormalization-group eigenvalues $y_t^{\rm IM} = 1$ and $y_h^{\rm IM} = 15/8$ for the case of the two-dimensional Ising model~\cite{Henkel}. 
These exponents govern the critical behaviour along the second-order transition line of the phase boundary and the associated critical exponents fall into the Ising universality class, where
\begin{equation}
 e_L  \sim L^{-1},  \quad
        c_L  \sim  L^{0}, \quad
        m_L  \sim L^{-1/8}, \quad
        \chi_L  \sim  L^{7/4}, \quad
        m_{T_{L}} \sim  L^{7/8}.\label{eq-IMuc} 
\end{equation}
In the above equation only the last two observables diverge with the system size in the vicinity of the transition and are then adapted for a numerical study, in particular for the determination of the pseudocritical point, since otherwise a possible regular contribution (\emph{e.g.}, in the case of the energy and of the specific heat) may hide partially the singular signal.
\begin{table}[H] \small
\begin{center}
\begin{tabular}{|| l | l l l 
|| l ||} 
 \multicolumn{4}{l}{\textbf{Ising model universality class} (FSS at $T_L$)} & \multicolumn{1}{c}{FSS at $H_L$} \\
 \hline
    Quantities &  $\chi_{12}\sim L^{y_t^{\rm IM}+y_h^{\rm IM}-D}$ & $\chi\sim L^{2y_h^{\rm IM}-D}$ & $ m_{T}\sim L^{y_t^{\rm IM}+y_h^{\rm IM}-D}$ & $ m_{T}\sim L^{y_t^{\rm IM}+y_h^{\rm IM}-D}$ 
    \\ [0.5ex] 
 Expected exponent &   7/8 (0.875) & 7/4 (1.75) & 7/8 (0.875) &  7/8 (0.875)
 \\ 
 \hline
 \hline
  $L=5-8$ &  0.867 (11) &  1.773 (8)  &  0.913 (7) 
    & 0.956 (7) \\ 
 \hline\hline
 $L=5-64$ &  0.8716 (5)  & 1.7617 (4) & 0.8806 (4)
 & 0.8837(3)\\ 
 \hline
 \hline
  $L=16-64$ &  0.8725 (8) &  1.7592 (6)  & 0.8763 (6) 
 & 0.8754(4) \\
 \hline
\end{tabular}
 \caption{Summary of critical exponents for the two-dimensional Blume-Capel model approaching the second-order zero crystal field transition point along the line $\Delta_{\rm c} = 0$. Data produced via the Wang-Landau ($L = 5 - 8$) and hybrid Metropolis-Wolff ($L > 8$) schemes.}
 \label{table:1}
 \end{center}
\end{table}
Along the second-order transition line the scaling field $g = \Delta - \Delta_{\rm c}$ associated to the crystal field (here $\Delta_{\rm c} = 0$, but it would take a non-zero value along the transition line) is a supplementary parameter which can also be used in the numerical analysis as it allows to prescribe other useful response functions, such as 
\begin{equation}
\label{chi_12}
\chi_{12}=\partial^2f_{\rm sing}(t,h,g)/\partial h\partial g, \quad
\chi_{2}=\partial^2f_{\rm sing}(t,h,g)/\partial g^2;
\end{equation}
see also equations~(\ref{eq-chi12}) and (\ref{eq-chi2}).

As discussed in Section~\ref{sec:BC_model} the phase diagram of the Blume-Capel is well established. Our objective here is to study first the convergence of the exponents to the Ising universality class. The next subsection will be devoted to the tricritical Ising model universality class. Simulations were therefore carried out at $T_{\rm c} = 1.6929$ ($\Delta_{\rm c} = 0$)~\cite{malakis10} where histogram reweighting was performed in the parameters $T$ or $h$. 

Figure~\ref{fig:comparison} displays the typical shape of some thermodynamic averages. In particular, the magnetocaloric coefficient and the fourth-order magnetisation cumulant for small system sizes $(L = 5 - 8)$ were obtained by temperature reweighting, comparing the two algorithmic approaches used. In a finite system, thermodynamic quantities do not present any singularity; fluctuations are limited by the size of the system. 
Thus, the maxima of these quantities lie close to the critical temperature in the thermodynamic limit  (vertical line exactly at $T_{\rm c} = 1.6929$). This temperature of the maxima, $T_L$, can be identified as the pseudocritical temperature of the corresponding quantity. Pseudocriticality may also be characterised in other directions of the parameters space, \emph{e.g.}, the magnetic-field direction, as the value $\pm H_L$ of the magnetic field for which diverging quantities are maximised.  

Further simulations were then performed up to sizes $L = 64$, which is still something fairly easy to achieve. Then, we proceeded to the ordinary FSS analysis, fitting the quantities to simple power laws in order to test equation~(\ref{eq-IMuc}).
Since we do not take corrections-to-scaling into account, 
the fits are linear. However, it is clear that finite-size effects are present. In order to get a qualitative imprint of their presence while approaching the asymptotic limit, we performed for each quantity of interest fits within three intervals of system sizes: (i) The small lattice-size regime ($L = 5 - 8$), (ii) the complete available lattice-size regime ($L = 5 - 64$), and (iii) the large lattice-size regime ($L = 16 - 64$).

\begin{table}\small 
\begin{center}
\begin{tabular}{|| l | l l l 
||} 
 \multicolumn{4}{l}{\textbf{Ising model universality class} (the role of corrections-to-scaling)} \\
 \hline
     Quantities &  $\chi_{12}\sim L^{y_t^{\rm IM}+y_h^{\rm IM}-D}$ & $\chi\sim L^{2y_h^{\rm IM}-D}$ & $ m_{T}\sim L^{y_t^{\rm IM}+y_h^{\rm IM}-D}$ 
    \\ [0.5ex]  
 Expected exponent &  7/8 (0.875) & 7/4 (1.75) & 7/8 (0.875)
 \\ 
 \hline
 \hline
  $L=5-8$ &  0.878(83) &  1.82(7)  &  0.94(6) 
  \\ 
 \hline\hline
 $L=5-64$    & 0.8727(9) & 1.7586(7) & 0.8745(7) 
  \\ 
 \hline
 \hline
  $L=16-64$ &  0.875(3) & 1.758(2)  & 0.877(2) 
  \\ 
 \hline
\end{tabular}
 \caption{Similar to table~\ref{table:1} but this time including corrections-to-scaling $(L^{-\omega})$ terms in the fits.}
 \label{table:2}
 \end{center}
\end{table}

As expected, the fits taking only the largest sizes into account provided the best estimates for the critical exponents -- the expected values are presented in table~\ref{table:1} for comparison. We note that fits with only the smallest sizes systematically gave results that are
not very accurate and we then proceeded to consider possible improvements with the inclusion of corrections-to-scaling terms. It should be noted however, that these effects are counterbalanced by the large sizes when all the sizes of the system are taken into account. The best results are those stemming from the magnetocaloric coefficient and susceptibility. The study by Zierenbeg \emph{et al.}~\cite{Zierenberg} reported a value of $\gamma/\nu=1.750 (3)$ for sizes within the range $L = 32 - L=128$, while Malakis \emph{et al.}~\cite{Malakis}, using Wang-Landau simulations and sizes $L = 50 - 100$ reported $\gamma/\nu=1.748(11)$ at the crystal field value $\Delta = 1$ (still in the second-order regime of the phase boundary where the model is governed by the Ising universality class). 
These results suggest that in our case scaling-correction effects appear for such sizes, even for the case $L\geq 16$. One can thus try to improve the fits by considering correction-to-scaling terms via the exponent $\omega$, which is known to take the value $\omega = 1.75$ for the two-dimensional Ising universality class~\cite{CorrectionIsing,sandvik16}. However, this strategy appeared to only slightly improve the extrapolations, when sizes up to $64$ were taken into account, but failed at improving small sizes estimates; see table~\ref{table:2} for more details.

\begin{table}[H] \small
\begin{center}
\begin{tabular}{|| l | l  l l l 
||} 
 \multicolumn{4}{l}{\textbf{Tricritical Ising model universality class} (FSS at $T_L$)} \\
 \hline
    Quantities & 
    $\chi_2\sim L^{2y_t^{\rm tri}-D}$  & $\chi_{12}\sim L^{y_t^{\rm tri}+y_h^{\rm tri}-D}$ & $\chi\sim L^{2y_h^{\rm tri}-D}$ & $ m_{T}\sim L^{y_t^{\rm tri}+y_h^{\rm tri}-D}$ 
    \\ [0.5ex] 
  Expected exponent &  
  8/5 (1.60) & 69/40 (1.725)  & 37/20 (1.85) & 69/40 (1.725) 
  \\ 
 \hline\hline
  $L=5-8$ &  
  1.488 (6) & 1.619(6)  & 1.763(5) & 1.709(5) 
 \\  
 \hline\hline
 $L=5-64$ &  
 1.6308(7)  & 1.7542(6) & 1.8808(6) & 1.7203(6) 
 \\ 
 \hline\hline

$L=16-64$ & 
1.657(3)  & 1.776(3) &  1.894(2)  &  1.748(2)
 \\ 

 \hline
\end{tabular}
 \caption{Summary of critical exponents for the two-dimensional Blume-Capel model at the tricritical point from FSS at the pseudocritical temperatures. The numerical approaches used for the different size spectrum are analogous to those described in table~\ref{table:1}.}
 \label{table:3}
 \end{center}
\end{table} 

\subsection{Scaling hypothesis and finite-size scaling at the tricritical point}

In the case of the Blume-Capel model, in order to describe  the approach to the tricritical point an additional field enters the description and it is more appropriate to introduce the scaling fields $t$ and $g$ as~\cite{Lawrie}
 
\begin{eqnarray}
   &&t= T-T_t,\\
   &&g=(\Delta-\Delta_{t})+at,\\
   && h=H
\end{eqnarray}
with the temperature $T$, crystal field $\Delta$, magnetic field $H$, and a coefficient $a$.
The third field $h = H$ describes the deviations from the tricritical point outside the symmetry plane, while $g$ and $t$ describe the approach within the symmetry plane. The free-energy density is then expressed as
\begin{equation}
     f_{\rm sing}(t,\Delta,h)  =L^{-d} f_{\rm sing}(L^{y_t}t, aL^{y_t}t(1+L^{-y_t+y_g}gt^{-1}), L^{y_h}h)
\end{equation}
It is more convenient in our case to work with the variable $\Delta$ rather than $g$ since it appears directly in the Hamiltonian.
At the tricritical point, critical exponents have been first conjectured from the dilute Potts model~\cite{Nijs, Nienhuis2, Pearson} and determined by conformal field theory~\cite{Cardy, Lassig,Henkel} as $y_t^{\rm tri} = 9/5$, $y_h^{\rm tri} = 77/40$, $y_g^{\rm tri} = 4/5$~\cite{Nienhuis}. 
The expected FSS forms in this case read as
\begin{equation}
        c_L  \sim  L^{8/5}, \quad
        \chi_L  \sim  L^{37/20}, \quad
        m_{T_{L}} \sim  L^{69/40}.
\end{equation}
Since the crystal field $\Delta$ is a linear combination of the scaling fields $t$ and $g$, derivatives with 
respect to $\Delta$ will lead to the following corrections-to-scaling
\begin{eqnarray}
 && \chi_{2_L}   \sim L^{2y_t^{\rm tri}-D}(1+bL^{-y_t^{\rm tri}+y_g^{\rm tri}})\sim L^{8/5}(1+bL^{-1}), \\
 && \chi_{{12}_L}   \sim L^{y_h^{\rm tri}+y_t^{\rm tri}-D}(1+bL^{-y_t^{\rm tri}+y_g^{\rm tri}})\sim L^{69/40}(1+bL^{-1})
\end{eqnarray}
for the two responses which involve the crystal-field fluctuations.

The same procedure as the one carried out at the critical point is also presented below, this time at the tricritical point; see also figure~\ref{fig:phase_diagram}. 
\begin{table}[H] \small
\begin{center}
\begin{tabular}{|| l |  l l l l 
||} 
 \multicolumn{4}{l}{\textbf{Tricritical Ising model universality class} (FSS at $\Delta_L$)} \\
 \hline
 \hline
    Quantities & $\chi_2\sim L^{2y_t^{\rm tri}-D}$  & $\chi_{12}\sim L^{y_t^{\rm tri}+y_h^{\rm tri}-D}$ & $\chi\sim L^{2y_h^{\rm tri}-D}$ & $m_T\sim L^{y_t^{\rm tri}+y_h^{\rm tri}-D}$ 
    \\ [0.5ex]  
  Expected exponent & 8/5 (1.60) & 69/40 (1.725)  & 74/40 (1.85) & 69/40 (1.725)
  \\ 
 \hline\hline
  $L=5-8$ & 1.603(2)  & 1.710(2)  & 1.820(2) & 1.601(4)

 \\ 
 \hline\hline

 $L=5-64$ &  1.6057(4) & 1.7234(3) & 1.8422(3) &  1.7191(4)

 \\ 
 \hline\hline

$L=16-64$ & 1.608(2)  & 1.730(2) &   1.851(2) & 1.740(2) 
 \\ 
 \hline
\end{tabular}
 \caption{Similar to table~\ref{table:3}. Exponent estimates obtained from fits at the pseudocritical crystal fields.}
 \label{table:4}
 \end{center}
\end{table} 
The FSS analysis of the thermodynamic quantities under study leads to the exponents disclosed in tables~\ref{table:3}
and \ref{table:4}. In table~\ref{table:3} the analysis is based on the maxima of the diverging quantities when the histogram reweighting is performed along the temperature direction. Therefore the quantities are measured strictly at $\Delta_{\rm t}$ and at the corresponding pseudocritical temperatures $T_{L}$. In table~\ref{table:4} we follow the exact reversed strategy, sitting at the corresponding pseudocritical values of the crystal field $\Delta_{L}$.

\section{Zeros of the partition function}
\label{sec:zeros}

\subsection{Fisher, crystal-field, and Lee-Yang zeros}

An alternative perspective to the study of critical phenomena is through the zeros of the partition function. These are fundamental quantities of special interest since their FSS properties encode universal features of underlying phase transitions~\cite{Zuber83,WU}. They appear when a certain parameter of the partition function (temperature, magnetic, or crystal field, or associated fugacities) attains complex values. The approach was first developed by Lee and Yang~\cite{Yang,Lee} who studied the lattice gas model via the grand partition function as a polynomial with a complex fugacity. Fisher then tackled the problem via the zeros in the complex temperature field, where, with increasing system size, the position of the zeros would get closer and closer to the real axis and the critical temperature. 
This line of research  was proven to be quite successful for the study of criticality in the Ising model leading to high-accuracy results even on the basis of small-scale studies and has also been successfully applied to other spin models as well, such as the Potts model~\cite{Chen96,Kim00} and the Blume-Capel model considered here.

\begin{figure}[ht]
    \centering
\includegraphics[width=\textwidth]{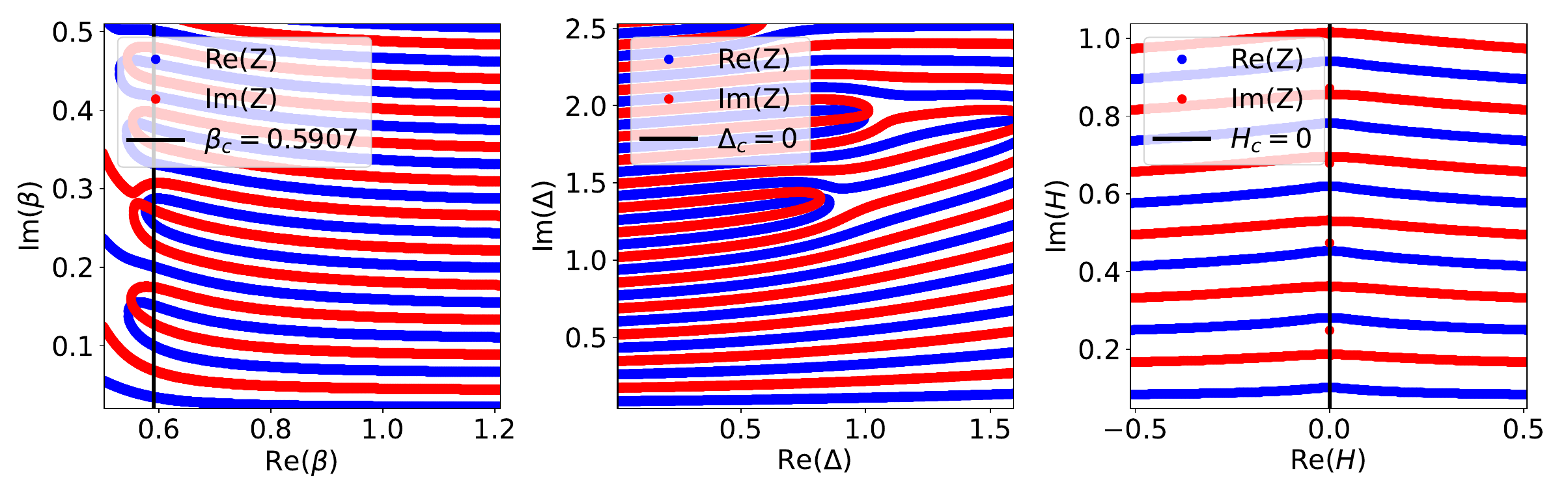}
    \caption{Fisher (\textbf{left panel}), crystal-field (\textbf{centre panel}), and Lee-Yang (\textbf{right panel}) zeros from Wang-Landau simulations for a system with linear dimension $L = 6$ in the vicinity of the critical point, see equation~(\ref{eq-points}). Note the inverse critical temperature $\beta_{\rm c} = 0.5907$ and that $H_{\rm c} = 0$. \label{fig:three-type_zeros}}
\end{figure}

Fisher zeros are the zeros in the complex temperature plane, while Lee-Yang zeros are those in the complex magnetic field plane. According to equation~(\ref{eq-split}), the partition function can be written for a complex temperature $\beta=\beta_r+i\beta_i$, as
\begin{equation}
{Z}(\beta_r, \beta_i) = \langle\cos[\beta_i (JE_J-\Delta E_\Delta)]\rangle_{\beta_r} + i\langle \sin[\beta_i (JE_J-\Delta E_\Delta)]\rangle_{\beta_r}.
\end{equation}
In the complex plane of the crystal field $\Delta=\Delta_r+\Delta_i$, 
the crystal-field partition function zeros will follow from the vanishing of
\begin{equation}
Z(\Delta_r, \Delta_i) = \langle\cos[\beta(-\Delta_iE_\Delta)]\rangle_{\Delta_r} + i\langle \sin[\beta(-\Delta_i E_\Delta)]\rangle_{\Delta_r}.
\end{equation}
\noindent
Eventually, in the complex $h$-plane corresponding to the external magnetic field $H$, we obtain the Lee-Yang partition function 
\begin{equation}
Z(H_r, H_i) =\langle\cos[\beta(H_i M)]\rangle_{H_r} + i\langle \sin[\beta(H_i M)]\rangle_{H_r}.
\end{equation}
It is important to note that for the present model under study the Lee-Yang theorem holds, \emph{i.e.}, all Lee-Yang zeros are purely imaginary, $H_{rn}=0$, and lie on a unit circle in the complex fugacity plane $e^{-2h}$. This theorem holds for all dimensionalities and/or lattice sizes. 

Figure~\ref{fig:three-type_zeros} illustrates the temperature (Fisher), crystal-field, and magnetic-field (Lee-Yang) zeros for the Blume-Capel model on a $6 \times 6$ square lattice in zero magnetic field at the critical point specified in equation~(\ref{eq-points}). The zeros are located in the area where the real part (shown in blue) and the imaginary part (shown in red) of the partition function are both zero for the same complex parameter ($T$, $\Delta$, or $H$), so that the partition function vanishes and therefore provides the position of a zero. 
\begin{table}[ht]
\begin{center}
\begin{tabular}{|| l | l | l  ||} 
 \multicolumn{3}{l}{\textbf{Ising model universality class}} \\
 \hline
    Zeros & $\operatorname{Im}(\beta_0)\sim L^{-y_t^{\rm IM}}$   & $\operatorname{Im}(H_0)\sim L^{-y_h^{\rm IM}}$  \\ [0.5ex] 
 Expected exponent & 1 &  15/8 (1.875) \\ 
 \hline
 \hline
  $L=5-8$ & 1.011(1)  &1.873(3)  \\ 
\hline\hline
 $L=5-64$ & 1.0000(3)   & 1.876(2) \\ 

 \hline\hline
 $L=16-64$ & 0.997(2)   &  1.878(9) \\

 \hline
\end{tabular}
 \caption{Summary of critical exponents $y_t^{\rm IM}$ (Fisher zeros) and $y_h^{\rm IM}$ (Lee-Yang zeros) for the two-dimensional Blume-Capel model at the critical point, of equation~(\ref{eq-points}), for $h_{\rm c} = 0$. Data produced via the Wang-Landau ($L = 5 - 8$) and hybrid Metropolis-Wolff ($L > 8$) schemes.}
 \label{table:5}
 \end{center}
 \end{table}
\begin{figure}[ht]
    \centering
\includegraphics[width=\textwidth]{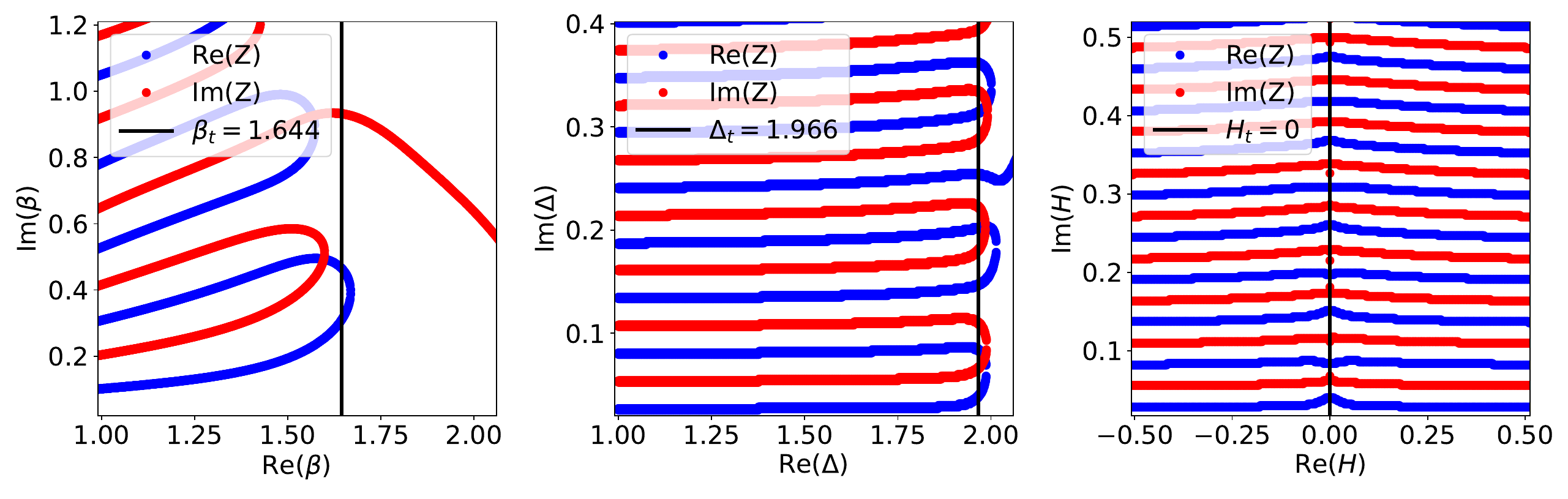}
\caption{Similar to figure~\ref{fig:comparison} but for the tricritical point. Note the inverse tricritical temperature $\beta_{\rm t} =  1.644$ and that $h_{\rm t} = 0$.
}\label{fig:zeros_tricritical}
\end{figure}
The figure shows only the changes in the sign of $\operatorname{Re}(Z)$ and $\operatorname{Im}(Z)$ from which we can deduce the value of the zeros of the partition function. As it is well-known, the FSS of the Fisher and Lee-Yang zeros gives access to the thermal exponent, $y_t^{\rm IM}$, and the magnetic exponent, $y_h^{\rm IM}$, respectively 
\begin{eqnarray}
       & \operatorname{Im}\beta_{n}(L)  \sim  L^{-y_t^{\rm IM}}, \quad  & \operatorname{Re} |\beta_{n}(L)-\beta_c|  \sim  L^{-y_t^{\rm IM}}, 
       \\
    & \operatorname{Im} H_{n}(L)  \sim  L^{-y_h^{\rm IM}}. 
\end{eqnarray}

Following our previous practice in the framework of FSS analysis, we carried out power-law fits over different size intervals. At this point we should note that we were only interested in the imaginary part of these zeros; of course the real part can also be used, but as the results were less interesting we excluded them for the needs of this study. For each type of zero we focused on the first zero $\beta_{i0}$ or $H_{i0}$, \emph{i.e.}, the zero closest to the real axis. In particular, one of the best estimates that we retrieved refers to the value $y_h^{\rm IM} = 1.876(2)$, which is in excellent agreement with the expected value $y_h^{\rm IM}={1.875}$. Inspecting table~\ref{table:5} that summarises the results from all our fitting attempts, it becomes evident that both Fisher and Lee-Yang zeros give critical exponents very close to the expected values for each case. What is more, the estimated exponents are remarkably accurate even when considering the FSS analysis over the small lattice-size regime, reflecting the fact that within this approach, finite-size corrections are negligible. This is in sharp contrast to the case of other thermodynamic averages considered before. Note that we have not shown here any fitting analysis based on the zeros of the crystal field at the critical point (only at the tricritical point; see below) as the results for the fits were found to be of low quality.

Next, we contemplate the case of the tricritical point. Graphically the zeros are shown in figure~\ref{fig:zeros_tricritical}. 
Compared with the critical point at the same size, the Fisher zeros are much less dense and the first zero has a much larger imaginary component. On the other hand, the crystal field and Lee-Yang zeros are very dense, even at small sizes. These first zeros almost form a vertical line. In fact, the crystal-field zeros have a similar behaviour to that of the Fisher zeros for the critical point, and vice versa. 
\begin{table}[H]
\begin{center}
\begin{tabular}{|| c c  c||} 
 \multicolumn{3}{l}{\bf Tricritical Ising model universality class} \\
 \hline
    Zeros &  $\operatorname{Im}(\Delta_0)\sim L^{-y_{t}^{\rm tri}}$  & $\operatorname{Im}(H_0)\sim L^{-y_{h}^{\rm tri}}$ \\ [0.5ex] 
 Expected exponent  & 9/5 (1.8) &  77/40 (1.925) \\ 
 \hline
 \hline 
  $L=5-8$  &  1.802(4)& 1.927(3)\\ 
 \hline\hline
 $L=5-64$  & 1.8077(7)&  1.924(2)\\ 
 \hline
 \hline
  $L=16-64$ & 1.803(9)&  1.922(5)\\ 
 \hline
 \hline
\end{tabular}
 \caption{Similar to table~\ref{table:5} but this time critical exponents were obtained from crystal-field and Lee-Yang zeros at the tricritical point.} 
 \label{tab:6}
 \end{center}
\end{table} 
\noindent 
In table~\ref{tab:6}, we summarise the results from the FSS analysis of the imaginary parts of the zeros in the complex plane corresponding to the temperature, crystal and magnetic fields, according to
\begin{eqnarray}
       & \operatorname{Im}\beta_{n}(L)  \sim  L^{-y_t^{\rm tri}}, \quad  & \operatorname{Re} |\beta_{n}(L)-\beta_t|  \sim  L^{-y_t^{\rm tri}}, 
       \\
    & \operatorname{Im}\beta_{n}(L)  \sim  L^{-y_t^{\rm tri}}, \quad  & \operatorname{Re} |\Delta_{n}(L)-\Delta_t|  \sim  L^{-y_t^{\rm tri}} , \\
    & \operatorname{Im} H_{n}(L)  \sim  L^{-y_h^{\rm tri}}.
\end{eqnarray}

We only focused on the crystal-field and Lee-Yang zeros since corrections were found to be very strong for the Fisher zeros and it was not possible to obtain a good exponent from these, unless a large proportion of the system sizes was discarded (as an example we quote the value $y_t^{\rm tri}=1.72(1)$ for fits within the range $L = 16 - 64$ of system sizes). The opposite is not true for the crystal-field and Lee-Yang zeros. In particular, at small sizes the exponents found were already those expected. As at the critical point, these zeros indicate that finite-size corrections are negligible. Nevertheless, we also did test power-law fits with corrections which improved the exponent too. We would like to remind the reader here that the zeros of the two-dimensional Blume-Capel model have already been studied in the complex temperature plane. Ayat and Care~\cite{Ayat} tackled the problem along the fugacity complex plane $u = e^{-\beta \Delta}$ and the temperature complex plane $t = e^{-\beta}$ and reported for a system with linear size $L = 22$ the values $y_t^{\rm tri}=1.814$ and $y_t^{\rm tri}=1.782$, respectively. 
Kim and Kwak~\cite{KimPF} also studied the zeros of the partition function in the complex temperature plane using Wang-Landau simulations and suggested the tricritical exponent $y_t^{\rm tri} = 1.83$. 

Although the conventional approach to probe efficiently the critical behaviour of spin models using numerical simulations of finite sizes $L$ dictates the need for $L\rightarrow \infty$, the results presented in this section advocate a different scenario: the critical properties of the Blume-Capel model may be accurately established from small-scale numerical simulations after employing the FSS analysis of the partition function zeros.
Other studies focusing on the zeros of the partition function for small system sizes have shown a similar trend, \emph{i.e.}, minor finite-size corrections. Using the cumulant method, Deger \textit{et al}.~\cite{Deger_2020ISING,Deger_2019} determined the critical exponents of the Ising model in two and three dimensions with good accuracy via the Lee-Yang and Fisher zeros using sizes up to $L = 15$ in two dimensions and $L = 10$ in three dimensions, respectively. The same authors~\cite{Deger_2018} also studied the phase transition in a molecular zipper with relatively small lengths up to $N=30$, establishing the expected exponents. The zeros were also suggested as an alternative route to probe logarithmic corrections in statistical physics. In fact, Kenna and Lang~\cite{Kenna_1993} have been able to numerically extract the logarithmic corrections, which were analytically predicted by the renormalization-group theory for the $\Phi^4$ model in four dimensions, using systems of linear sizes $L \leq 24$. Additionally, through the density of zeros one may gain important information and determine the order and strength of a transition, as Janke and Kenna~\cite{janke2000strength} have shown for several systems, again using relatively small system sizes. Interestingly, the study by Alves \emph{et al.}~\cite{Alves_2002} contrasted the Janke and Kenna approach~\cite{janke2000strength} with another classification scheme proposed by Borrmann \emph{et al.}~\cite{Borrmann_2000} which explores the linear behavior for the limiting density of zeros proposed, and suggested that it is indeed possible to characterize the order of the transition in the four- and five-states Potts model via a small-size scaling analysis. On the other hand, there are no real explanations or hypotheses provided in the vast literature around this topic. One suggestion that we bring forward here, in support of the above observations that the finite-size corrections to the zeros of the partition function are small, is that the zeros may not comprise of regular contributions, or at least are strongly dominated by the singular part.

\subsection{Impact angle and amplitude ratios}

In the complex plane of the temperature, the zeros in the vicinity of the critical point $\beta_{\rm c}$ follow a line which should pinch the real axis at the critical temperature $T_{\rm c}$. 
\begin{table}[ht]
\begin{center}
\begin{tabular}{||  c c c c  c c c ||} 
 \multicolumn{7}{l}{\bf Tricritical Ising model universality class} \\
 \hline
 $L$   & $\varphi_{1,2}({\Delta})$  &  $\varphi_{1,2}(u)$  &  $\varphi_{2,3}(\Delta)$  &  $\varphi_{2,3}(u)$ & $\varphi_{1,3}(\Delta)$ &  $\varphi_{1,3}(u)$  \\ [0.5ex] 
 \hline\hline
\hline
 8  &  89.47 & 84.72 & 82.77 & 88.26 & 84.32 & 88.89 \\
 32  & 89.82 & 89.79 & 89.94 & 89.32& 89.67 &  89.77 \\
 56  & 89.54 & 89.68 & 89.81 &  89.92& 89.67 & 89.87 \\
 64  & 89.09 & 89.21  & 89.86 & 89.93  & 89.46 & 89.62\\
 \hline 
\end{tabular}
 \caption{Estimation of $\varphi$ at different system sizes and angles (first, second, and third zeros) for the two-dimensional Blume-Capel model at the tricritical point. Expected value: $\varphi_{2D}=90\degree$.}
 \label{tab:7}
 \end{center}
\end{table}
The real axis of $\beta$ and this line form an impact angle $\varphi$ which is related to the exponent $\alpha$ and the amplitude ratio of the specific heat via the relation                                       
\begin{equation}
\label{eq:phi}
    \tan{[(2-\alpha)\varphi]}=\frac{\cos(\pi \alpha) - A_+/A_-}{\sin(\pi \alpha)},
\end{equation}
if one defines the impact angle $\varphi$ between the positive sense of the axis and the zeros. The critical exponent $\alpha$ and the amplitude ratio $A_+/A_-$ are universal quantities, thus the impact angle is itself universal. For the two-dimensional Ising universality class the estimate of the amplitude ratio $A_+/A_- = 1$ from Monte Carlo simulations ~\cite{Delfino_1998} suggests that $\varphi_{2D}=\pi/2$. The values are the same also for the $\phi^6$ model~\cite{Fioravanti}. The impact angle is usually attained in the complex plane of the temperature, and we show here that it is also possible to extract this angle, and therefore the amplitude ratio $A_+/A_-$, in the crystal-field plane at the tricritical point. Thus, at the tricritical point the study was undertaken at the complex crystal-field plane and the exponential plane $u = e^{-\beta \Delta}$. There are several ways to retrieve the impact angle. We elaborate here on three different ways for each system size: (i) the angle between the first zero and the real crystal field, (ii) the angle between the second zero and the real critical crystal field, and (iii) the angle between the first and second zero with respect to the real axis. Table~\ref{tab:7} recaps the results for system sizes up to $L = 64$.

The angle $\varphi_{j,k}$ is the angle between the real $\Delta$-axis and the line from $\Delta^{(j)}(L)$ and $\Delta^{(k)}(L)$. Undoubtedly, for sizes in the range $L = 32 - 64$ all the angles are asymptotically close to the expected value $\varphi_{2D} = 90\degree$, allowing us to safely extract an estimate for the average angle $\varphi= 89.68(9)$, taken as an average over the $u$-plane estimates. From that one may extract the value $A_+/A_- = 1.07(6)$ for the amplitude ratio using equation~(\ref{eq:phi}) and the exact value of $\alpha$. To the best of our knowledge this is the first numerical estimation of the tricritical amplitude ratio in the two-dimensional Blume-Capel model.

\section{Partition function zeros: Energy and magnetisation cumulants}
\label{sec:cumulants}

Another powerful method to extract the partition function zeros from fluctuations of thermodynamic observables, such as the energy and the magnetisation, has been developed by Deger \emph{et al.}~\cite{PhysRevLett.110.050601,Deger_2018,Deger_2019,Deger_2020WEISS}. The advantage of the so-called cumulant method is that it does not require the computation of the full density of states, only usual simulations at fixed external parameters. The cumulant method has already been successfully tested to extract the Fisher and Lee-Yang zeros of several models, including a molecular zipper~\cite{Deger_2018}, the Ising model at two and three dimensions~\cite{Deger_2020ISING,Deger_2019}, and the Curie-Weiss model~\cite{Deger_2020WEISS}. The goal in this section is to test the cumulant method on the basis of the Lee-Yang zeros at the critical and tricritical points of the two-dimensional Blume-Capel model and compare the outcomes with our previous results. 

\subsection{From cumulants to zeros}

The partition function $\mathcal{Z}(q)$ contains fluctuations of thermodynamic quantities which are determined by the derivatives of the free energy with respect to a control parameter $q$ of a thermodynamic observable $\Phi$

\begin{equation}
        \langle \langle \Phi^n (q)\rangle \rangle = (-1)^n \partial_{q}^n \ln \mathcal{Z}(q)/N. \\ 
\end{equation}
For instance, for the total energy one has $\Phi \equiv E$ and if the control parameter is the inverse temperature, then
\begin{equation}
        \langle \langle E^n (\beta)\rangle \rangle = (-1)^n \partial_{\beta}^n \ln \mathcal{Z}/N, \\ 
\end{equation}
and the energy cumulants are generally expressed as non-linear combinations of central moments 
\begin{equation}
\mu^{n}= \langle (E - \langle E\rangle)^{n} \rangle / N. 
\end{equation}

The first cumulant gives the mean while the second cumulant is the variance, related respectively to the average energy and specific heat. One then has, up to $n=4$ 
\begin{equation}
        \kappa^{1} = \langle E \rangle/N, \quad 
        \kappa^{2} = \mu^{(2)}, \quad
        \kappa^{3} = \mu^{(3)}, \quad
        \kappa^{4} = \mu^{(4)} - 3N \mu^{(2)^2},     
\end{equation}
where $\langle \langle E^n \rangle \rangle  \equiv \kappa^{(n)}$. Note that one can find expressions of the cumulants up to the tenth order in reference~\cite{Janke_1997}.
The relation between the cumulants and the partition function zeros is given by 
\begin{equation}
    \langle \langle \Phi^n(q) \rangle \rangle = -(n-1)! \sum_k (q_k-q)^{-n}/N,
\end{equation}
where $q_k$ are the zeros of the thermodynamic observable in the complex plane of the control parameter $q$. High-order cumulants encapsulate precious information about zeros, especially the first zero, \emph{i.e}, the one closest to the real axis, since it dominates the sum. The contributions from subleading zeros are suppressed with the cumulant order $n$~\cite{Deger_2018}. 

The phase transition of the Blume-Capel model depends on three control parameters: (i) the inverse temperature $q =-\beta$, (ii) the reduced external magnetic field $q=\beta H$, and the reduced crystal field $q=\beta \Delta$. Thus: \\
(1) For the case of the energy $\Phi \equiv E$ and its conjugate variuable $q=-\beta$ one obtains the Fisher zeros.\\
(2) For the case of the magnetisation $\Phi \equiv M$, and its conjugate variable $q=\beta H$ one may extract the Lee-Yang zeros.\\
(3) We also expect that the quantity $E_\Delta$ and its conjugate parameter $q=-\beta \Delta$ give access to the zeros of the crystal field. \\
Also, different studies~\cite{PhysRevLett.110.050601,Deger_2018,Deger_2019,Deger_2020WEISS} show that one can extract the leading partition function zeros from
\begin{equation} 
     \operatorname{Re}[q_0-q]  \approx \frac{n(n+1) \langle \langle \Phi^n(q) \rangle \rangle \langle \langle \Phi^{n+1}(q) \rangle \rangle - n(n-1) \langle \langle \Phi^{n-1}(q) \rangle \rangle \langle \langle \Phi^{n+2}(q) \rangle \rangle}   {2 \left[(n+1) \langle \langle \Phi^{n+1}(q) \rangle \rangle^2 - n \langle \langle \Phi^n(q) \rangle \rangle  \langle \langle \Phi^{n+2}(q) \rangle \rangle \right]} 
     \end{equation}
\begin{equation} 
    |q_0-q|^2  \approx  \frac{n^2(n+1) \langle \langle \Phi^n(q) \rangle \rangle^2  - n(n^2-1) \langle \langle \Phi^{n-1}(q) \rangle \rangle \langle \langle \Phi^{n+1}(q) \rangle \rangle}   { (n+1) \langle \langle \Phi^{n+1}(q) \rangle \rangle^2 - n \langle \langle \Phi^n(q) \rangle \rangle  \langle \langle \Phi^{n+2}(q) \rangle \rangle } 
    \end{equation}
for $n\gg 1$.

\begin{figure}[H]
    \centering
        \begin{subfigure}[b]{0.49\textwidth}
     \includegraphics[width=\textwidth]{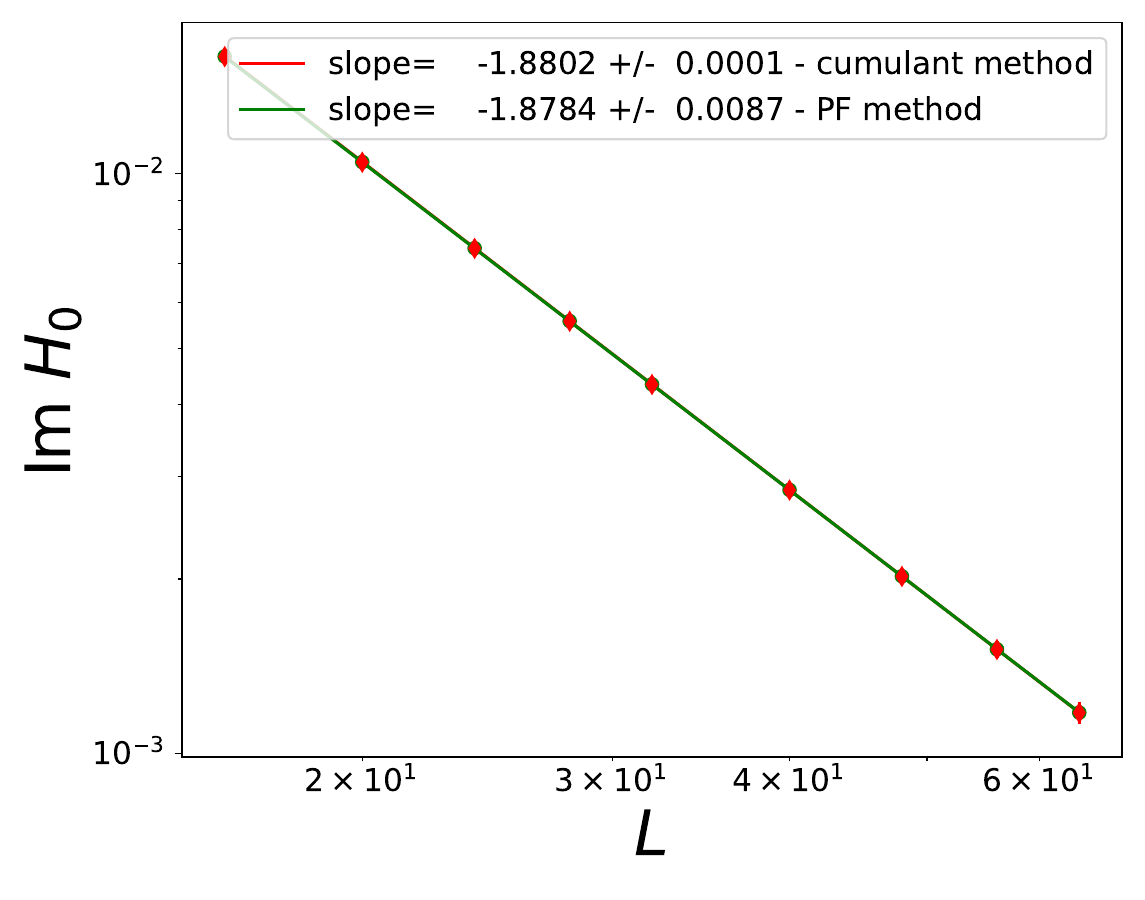}
    \end{subfigure}
    \begin{subfigure}[b]{0.49\textwidth}
    \includegraphics[width=\textwidth]{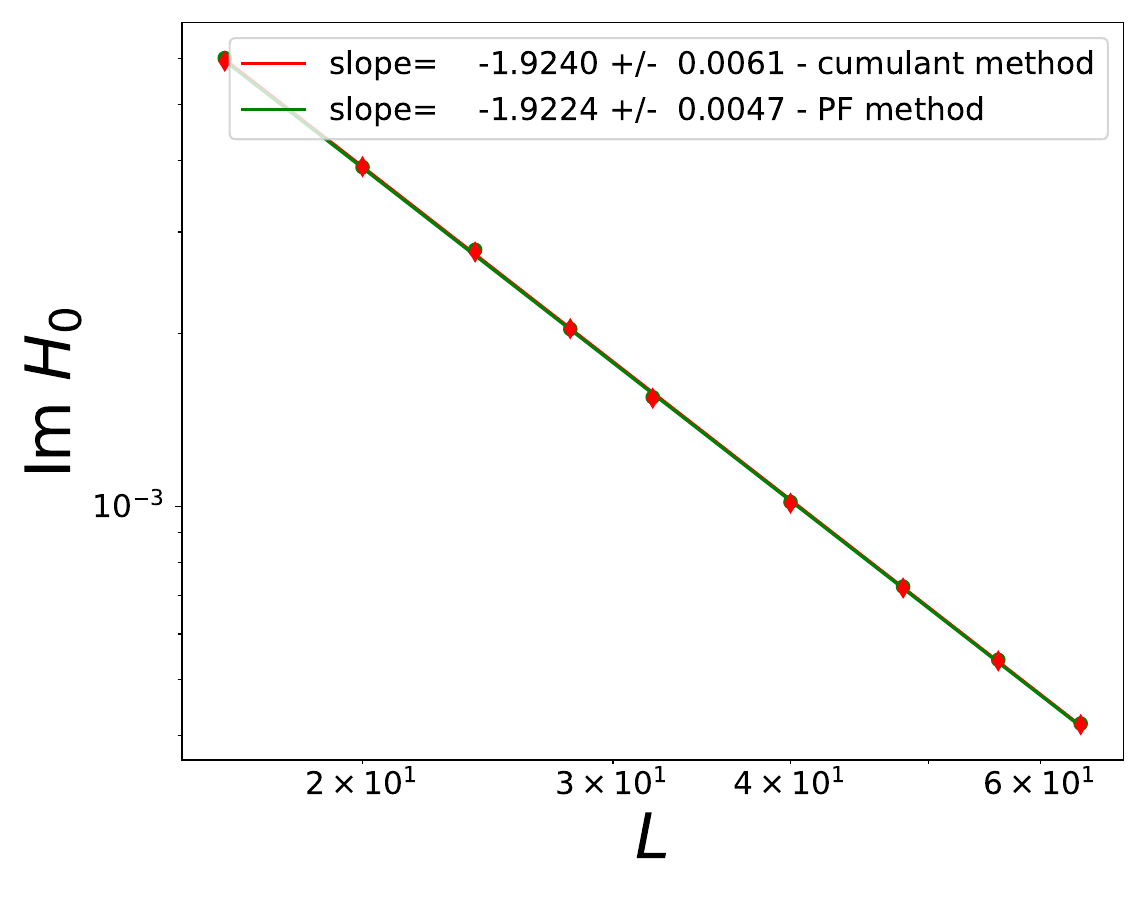}
    \end{subfigure}
\caption{FSS behaviour of the Lee-Yang zeros at the critical (\textbf{left panel}) and tricritical (\textbf{right panel}) points. Comparative results from the cumulant and  partition function (PF) methods are shown, via the hybrid Metropolis-Wolff algorithm. We quote the expected values $y_h^{\rm IM} = -1.875$ and $y_h^{\rm tri} = -1.925$  for the benefit of the reader. Lines are simple linear fits in a double-logarithmic scale.}
\label{fig:cumulant1}
\end{figure}

\subsection{The case of Lee-Yang zeros}

For the case $H=0$, odd cumulants $\langle \langle M^{2n+1} \rangle \rangle$ vanish due to the symmetry properties. Hence, equation~(37) simplifies to
\begin{equation}
    \operatorname{Im}[H_0] \approx \pm \frac{1}{\beta}\sqrt{ 2n(2n+1) \left|
    \frac{\langle \langle M^{2n} (0)\rangle \rangle }{\langle \langle M^{2(n+1)}(0) \rangle \rangle }\right|}.
\end{equation}
Monte Carlo simulations for systems with linear sizes $L=16 - 64$ were performed at the critical $(\Delta_{\rm c} = 0,\; T_{\rm c} = 1.6929)$ and tricritical $(\Delta_{\rm t} = 1.966, \; T_{\rm t} = 0.608)$ points for $n=3$. Figure~\ref{fig:cumulant1} compares the results from the cumulant method and those from the  partition function analysis secured in the previous section. The values of the zeros obtained by the two methods are very close, as one can inspect from  figure~\ref{fig:cumulant1} where the difference between the two fits is barely perceptible. At the critical point the estimate $y_h=1.8802(1)$ is computed in agreement with the expected value $y_h^{\rm IM}=1.875$, while at the tricritical point the exponent value found $y_h=1.9240(61)$ is in perfect agreement with the expected result $y_h^{\rm tri}=1.925$. Therefore, we may safely conclude that the cumulant method is a very efficient way to extract the exponent $y_h$, both at the critical and tricritical points, involving at best the computation of the sixth- and eight-order cumulant of the magnetisation. 

Another interesting aspect worthy of investigation relates to the crossover phenomena at the tricritical point and the sensitivity of the Lee-Yang zeros on the transition. Let us now focus only on the tricritical point and explore its critical behaviour by  arbitrarily selecting two additional values of the temperature in close proximity to the tricritical value $T_{\rm t} = 0.608$. Namely one temperature below $T_{\rm t}$, $T = 0.6075$, and another one above $T_{\rm t}$, $T=0.6083$. The crystal field is fixed at its tricritical value $\Delta_{\rm t} = 1.966$~\cite{Kwak2015}. 
\begin{figure}[H]
  \centering
    \includegraphics[width=0.6\linewidth]{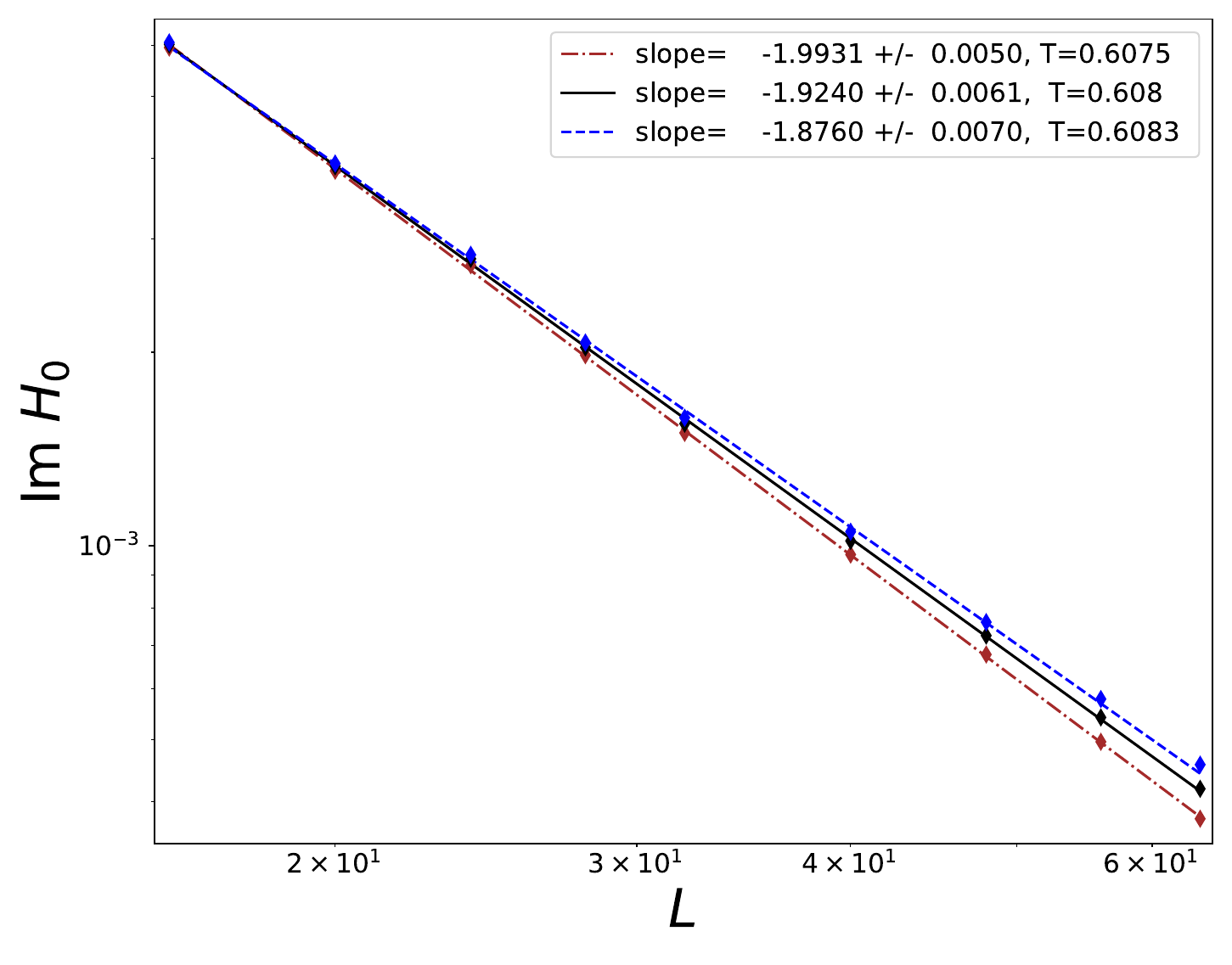}
    \caption{FSS behaviour of the Lee-Yang zeros at three different temperatures around the tricritical point, as indicated in the panel, extracted by the cumulant method. using the hybrid Metropolis-Wolff algorithm. Lines are simple linear fits in a double-logarithmic scale.}
\label{fig:cumulant2}
\end{figure}
Figure~\ref{fig:cumulant2} illustrates the fits for each temperature. Three distinct behaviours can be identified and several comments are in order at this point: (i) Below the tricritical temperature the exponents found quickly converge to the first-order characteristic exponent $y_h = D$. We therefore present evidence of the first-order phase transition. (ii) Above the tricritical temperature we see that the exponent found is remarkably close to that of the Ising ferromagnet, falling back into the critical behaviour of the second-order transition line and the Ising universality class. (iii) These results highlight that the Lee-Yang zeros at the tricritical point are extremely sensitive quantities and it is therefore extremely important to know precisely the location of the critical line and tricritical point of the model in order to extract securely the critical exponents. (iv) Finally, scanning the phase diagram from $T = 0.6075$ to $T = 0.6083$, it is possible to characterize all the richness of the different transitions of the two-dimensional Blume-Capel model and to extract the exponent $y_h$ in a remarkably accurate manner.

\subsection{Finite-size scaling of the cumulants}

\begin{figure}[H]
\centering
    \begin{subfigure}[b]{0.46\textwidth}        \includegraphics[width=\textwidth]{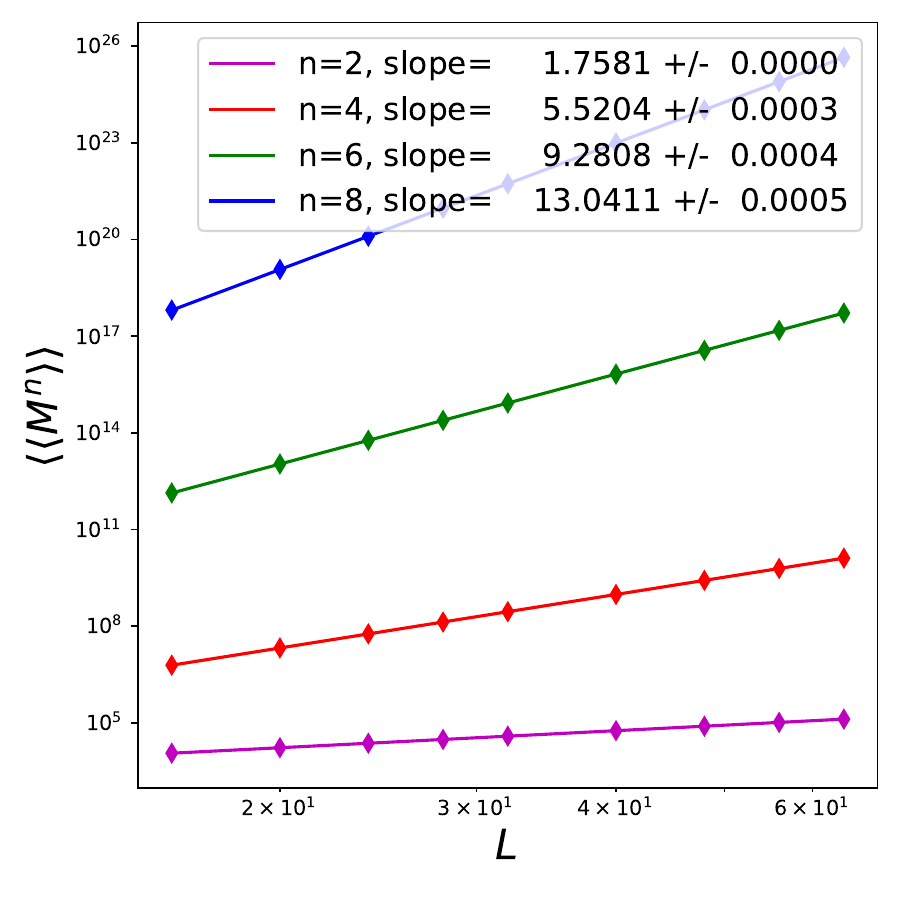}
    \end{subfigure}
       \begin{subfigure}[b]{0.46\textwidth}
    \includegraphics[width=\textwidth]{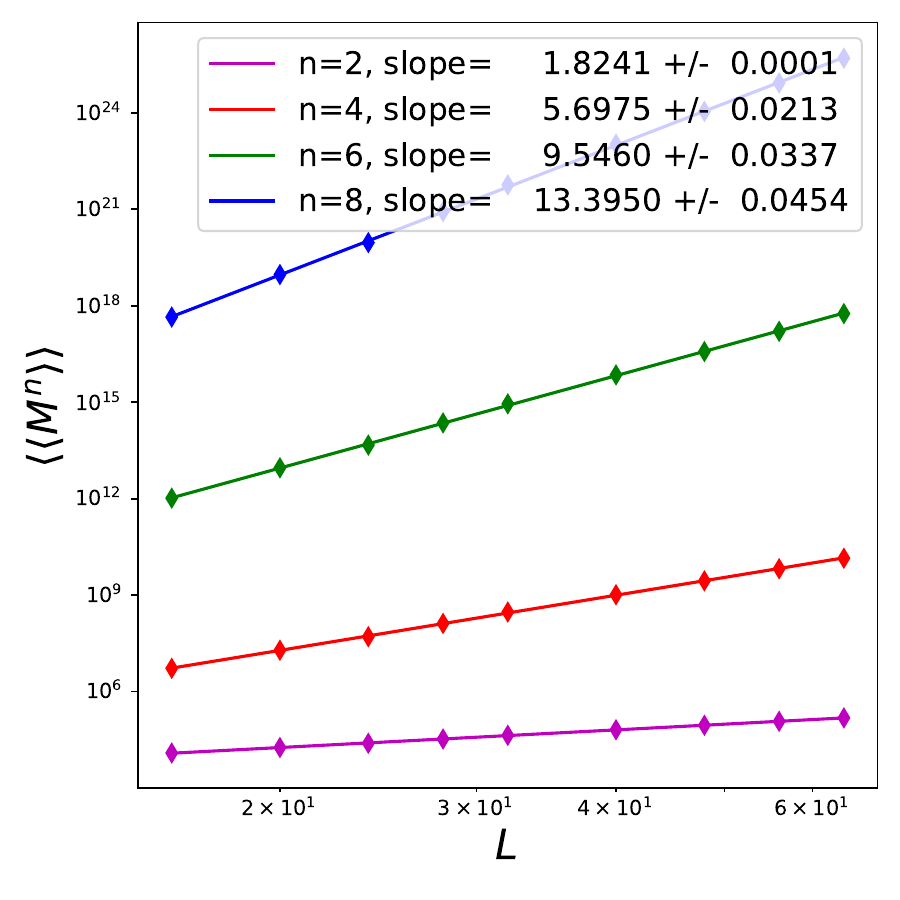}
    \end{subfigure}
\caption{FSS behaviour of the magnetisation cumulants $\langle \langle M^n \rangle \rangle$ at the critical (\textbf{left panel}) and tricritical (\textbf{right panel}) points according to the scaling laws $\langle \langle M^n \rangle \rangle \sim L^{-D+ny^{\rm IM}_h}$ and $\langle \langle M^n \rangle \rangle \sim L^{-D+ny^{\rm tri}_h}$, respectively. Numerical results for $n=2, 4, 6,$ and $8$ are shown, via the hybrid Metropolis-Wolff protocol. Note the double-logarithmic scale.}
    \label{fig:cumulant3}
\end{figure}

Once we have computed the cumulants, an ordinary FSS analysis is also welcome, and we expect the following behaviours
\begin{equation}
\langle \langle E^n \rangle \rangle \propto L^{-D+ny_t},
\end{equation}
\begin{equation}
\langle \langle M^n \rangle \rangle \propto L^{-D+ny_h}.
\end{equation} 
Figure~\ref{fig:cumulant3} shows as an example the power-law behaviour of the magnetisation cumulants  at the critical point. 
We only focus on the even cumulants since the odd ones vanish. As one can see, the exponents are really close to the expected ones, both at $T_{\rm c}$ and at $T_{\rm t}$ for the large lattice-size regime $L = 16 - 64$. At the critical point, we extract $\langle \langle M^2 \rangle \rangle \sim L^{1.7581}$, $\langle \langle M^4 \rangle \rangle \sim L^{5.5204 (3)}$, $\langle \langle M^6 \rangle \rangle \sim L^{9.2808 (4)}$, and $\langle \langle M^8 \rangle \rangle \sim L^{13.0411 (5)}$ where one expects the values $1.75$, $5.50$, $9.25$, and $13.00$, respectively. The exponent estimates are already excellent for the small lattice-size regime, except for the case $n=2$. Indeed, one finds  $\langle \langle M^2 \rangle \rangle \sim L^{1.5922}$, $\langle \langle M^4 \rangle \rangle \sim L^{5.500 (6)}$, $\langle \langle M^6 \rangle \rangle \sim L^{9.251 (9)}$, and $\langle \langle M^8 \rangle \rangle \sim L^{13.00(2)}$. This consists strong evidence that the convergence is really robust for these higher-order cumulants and is also in contradiction to the scaling of the susceptibility ($\chi \sim L^{1.661 (9)}$ instead of $\sim L^{1.75}$) and the magnetocaloric coefficient ($m_{T} \sim L^{0.956(7)}$ instead of $\sim L^{0.875}$).

At the tricritical point the expected values are  $1.80$, $5.704$, $9.55$, and $13.40$ respectively, which are again remarkably in agreement with our results for the lattice-size window $L = 16 - 64$: $\langle \langle M^2 \rangle \rangle \sim L^{1.8241(1)}$, $\langle \langle M^4 \rangle \rangle \sim L^{5.70 (3)}$, $\langle \langle M^6 \rangle \rangle \sim L^{9.55 (4)}$, and $\langle \langle M^8 \rangle \rangle \sim L^{13.39 (5)}$.Note that at the tricritical point the FSS behaviour of quantities such as the susceptibility and the magnetocaloric coefficient is also not as good ($\chi \sim L^{1.857(6)}$ instead of $\sim L^{1.85}$ and $m_{T}\sim L^{1.707(5)}$ instead of $\sim L^{1.725}$).

\section{Conclusions}
\label{sec:conclusions}

We have analysed the Fisher, crystal-field, and Lee-Yang zeros of the two-dimensional Blume-Capel model at the critical and tricritical points, employing Wang-Landau and hybrid simulations of Metropolis and Wolff type. The main conclusion of our work is that the zeros of the partition function allow us to find exponents that are remarkably close to their expected values even by the simulation of very small system sizes. When compared with the exponents extracted from basic thermodynamic quantities, we notice that the zeros give more accurate results, without the need to include any scaling corrections. At the tricritical point, the crystal-field zeros allow us to obtain a very good value for the thermal exponent, unlike the Fisher zeros, which require the inclusion of correction terms. At the critical point, the opposite situation occurs. Our work calls attention to the fact that the zeros are highly sensitive to the external parameters of the system, especially around the area of the tricritical point of the phase boundary where strong crossover effects are expected to obscure any attempt for finite-size scaling. An alternative technique which measures the zeros benefiting from high-order cumulants of the energy and the order parameter has also been implemented in the present work and it was shown that this method also leads to concrete results of similar numerical accuracy, in reduced computational times. Finally, we have extended the standard analysis of the impact angle of the accumulation of Fisher zeros in the complex temperature plane to the complex crystal-field plane,  the first numerical computation of this specific amplitude ratio at tricriticality. The extension to crystal-field zeros is also a novelty of the present work.

\section*{Acknowledgments}
The work of NGF was supported by the  Engineering and Physical Sciences Research Council (grant EP/X026116/1 is acknowledged). RK has expressed his intention to thank C~B Lang, as this work is a continuation of what they started 30 years ago. BB, YH, and LM would like to thank the  Doctoral College for the Statistical Physics of Complex Systems and the $L^4$ collaboration, especially the interesting discussions with Yulian Honchar, Marjana Krasnytska, and Andy Manapany.

\printbibliography

\end{document}